\documentclass[10pt,a4paper,onecolumn]{article}
\usepackage[utf8]{inputenc}
\usepackage{amsmath}
\usepackage{amsmath,amsthm,amssymb,mathrsfs}
\usepackage{mathtools}
\usepackage{graphicx}
\usepackage{authblk}
\usepackage{textcomp}

\usepackage{booktabs}
\usepackage{color}
\usepackage[margin=0.8in]{geometry}
\usepackage[colorlinks=true, urlcolor=black]{hyperref}
\begin{document}
\newcommand{\ed}[1]{\textcolor{black}{#1}}
\newcommand{\edrs}[1]{\textcolor{black}{#1}}
\newcommand{\edrstwo}[1]{\textcolor{black}{#1}}
\newcommand\blfootnote[1]{%
  \begingroup
  \renewcommand\thefootnote{}\footnote{#1}%
  \addtocounter{footnote}{-1}%
  \endgroup
}
\newcommand{\degc}{\,^{\circ}\mathrm{C}}
\newcommand{\norm}[1]{\| #1 \|}
\newcommand{\ud}{\mathrm{d}}
\newcommand{\oh}{\frac{1}{2}}
\newcommand{\mb}[1]{\mathbf{#1}}

\title{Active elastohydrodynamics of vesicles in narrow, blind constrictions}
%\author{T. G. Fai$, R. Kusters$^{\dagger}$, J. Harting, C. Rycroft, and L. Mahadevan}
\author[1,$\dagger$]{T. G. Fai}
\author[2,$\dagger$]{R. Kusters}
\author[2,4]{J. Harting}
\author[1]{C. H. Rycroft}
\author[1,3,$^*$]{L. Mahadevan}
 \affil[1]{John A. Paulson School of Engineering and Applied Sciences, Harvard University, Cambridge, Massachusetts 02138,USA}
\affil[2]{Department of Applied Physics, Eindhoven University of Technology,\\ Eindhoven, The Netherlands} 
\affil[3]{Department of Physics, Department of Organismic and Evolutionary Biology, Harvard University, Cambridge, Massachusetts 02138,USA }
\affil[4]{Helmholtz Institute Erlangen-Nürnberg for Renewable Energy (IEK-11) \\Forschungszentrum Jülich, 90429 Nürnberg, Germany} 
\affil[$\dagger$]{These authors made equal contributions}
\affil[$^*$]{Corresponding author, \url{lm@seas.harvard.edu}}
\maketitle

\begin{abstract}
 Fluid-resistance limited transport of vesicles through narrow constrictions is a recurring theme in many biological and engineering applications. Inspired by the motor-driven movement of soft membrane-bound vesicles into closed neuronal dendritic spines, here we study this problem using a combination of passive three-dimensional simulations and a simplified semi-analytical theory for active transport of vesicles that are forced through such constrictions by molecular motors. We show that the motion of these objects is characterized by two dimensionless quantities related to the geometry and the strength of forcing relative to the vesicle elasticity. We use numerical simulations to characterize the transit time for a vesicle forced by fluid pressure through a constriction in a channel, and find that relative to an open channel, transport into a blind end leads to the formation of a \edrstwo{smaller} effective lubrication layer that strongly impedes motion.  When the fluid pressure forcing is complemented by forces due to molecular motors that are responsible for vesicle trafficking into dendritic spines, we find that the competition between motor forcing and fluid drag results in multistable dynamics reminiscent of the real system. Our study highlights the role of non-local hydrodynamic effects in determining the kinetics of vesicular transport in constricted geometries.  
\end{abstract}

\section{Introduction}

%\blfootnote{$^{\dagger}$: Equally contributing authors}

Intracellular transportation of vesicles frequently involves translocation through channels and into narrow pockets. An example of such a  process that has received considerable experimental attention with the advent of live-cell imaging is the motor-driven transportation of protein-rich vesicular endosomes through the necks of dendritic spines in neurons \cite{Park2006, Wang2008, Da2015}. The micron-sized vesicles deform strongly during transport as they are squeezed through the sub-micron sized necks by molecular motors. These vesicles play a critical role in transport of chemicals responsible for the normal functioning of neuronal synapses, and understanding the biophysical basis for and limits to their transport is thus an important problem. Indeed, this situation occurs not only in intracellular trafficking, but also in numerous cellular and microfluidic settings in which elastic bodies, such as manufactured elastic capsules \cite{Dawson2015,Duncanson2015}, hydrogels \cite{Li2015} or living cells \cite{Gabriele2010, Byun2013, Bagnall2015} squeeze through narrow constrictions. While driving forces such as pressure gradients, fluid flow, molecular motors, or external fields may promote passage, the changes in shape required for passage result in energetic barriers to transportation, altering the passage time and eventually leading to non-passage of the vesicle. Such transport processes are thus determined by a subtle balance between forces originating from various physical sources and the geometry of the constriction.

\begin{figure}
\centering
\includegraphics[width=0.76\textwidth]{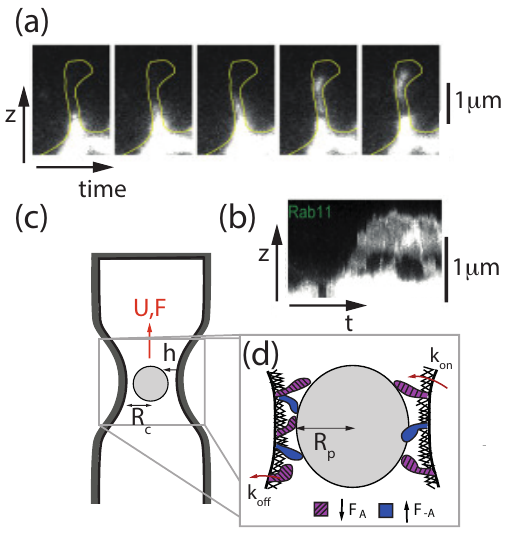}
\caption{(Color online) (a) Timelapse images and (b) corresponding kymograph from a dual-color timelapse recording showing the actomyosin-based transportation of an endosome through the neck of a neuronal dendritic spine over 40 -- 60 seconds (adapted from \cite{Da2015} and available under CC BY NC ND license). The bright color represents the endosome and the curve in (b) indicates the contour of the spine. The region of high intensity at the base of the spine represents a pool of endosomes, from which the kymograph shows one entrance event. (c) Our corresponding model system for the forced translocation of a vesicle through a narrow constriction. (d) Schematic of the motor model used which contain two species of motors, one pushing up the channel, and another pushing down the channel.  \label{figure1}}
\end{figure}

In recent years, considerable progress has been made in studying this process computationally in the context of the translocation of elastic capsules through cylindrical \cite{Queguiner1997, Park2013, Kusters2014, Rorai2015} and square \cite{Kuriakose2011, Moon2016} micro-channels. These studies typically consider transit through open channels (i.e. infinite channels or those with periodic boundary conditions at each end) in contrast with transportation into a closed constriction, i.e.\ one with a blind end, as occurs in dendritic spines (Fig.~\ref{figure1}(a) and (b)). Dendritic spines are bulbous structures protruding from dendrites. The heads of dendritic spines are the locations of postsynaptic densities, which are membrane receptor-enriched regions responsible for detecting neurotransmitters released into the synaptic cleft. Therefore a flux of membrane receptors to the spine head is required to maintain synaptic function, and these receptors are actively transported to dendritic spines on membrane-bound vesicles known as recyling endosomes. The nearly occlusive movements of these elastic vesicles through the narrow spine necks leads to large deformations of the vesicle. Moreover, as a consequence of the spine geometry, when the vesicle enters the constriction either passively due to an ambient pressure or due to the activity of motors, fluid enclosed by the spine must flow out of the same constriction in a narrow layer surrounding the vesicle. This leads to long-range interactions mediated by fluid incompressibility, a problem that is the particular focus of this study.

Here, we quantify how the translocation time through open and closed channels with a constriction depends on the constriction size, vesicle elasticity, and applied force. In Section \ref{sec:abs} we outline the abstraction of the experimental observations of endosomal transport. In Section \ref{sec:LB} we present a lattice Boltzmann method that we use to study the  translocation dynamics of a soft vesicle, focusing on how the applied force and channel geometry impact the transit speed of the vesicle (Fig.~\ref{figure1}(c)). In Section \ref{sec:lube} we describe a simplified model based on lubrication theory that captures the essential features of translocation determined using lattice Boltzmann simulations. In Section \ref{sec:lube_motor}, we apply the lubrication model to the phenomenon of intracellular cargo transportation by replacing the assumption of prescribed forcing by a realistic description of molecular motors, and in Section \ref{sec:compex} we compare the results of simulations to experimental observations on vesicle transportation into dendritic spines. We conclude with a discussion of how our model captures the diverse behaviors observed in vesicle translocation through spines, including vesicle rejection, corking, and translocation, and we show how different behaviors can be obtained by varying model parameters such as the constriction geometry and the dynamics of the molecular motors.

\section{Computational formulation of 3D endosomal transport}
\label{sec:abs}

Translocation of elastic vesicles through narrow channels is critical in the context of neuronal regulation associated with synaptic activity. The vesicles, also called endosomes, serve to transport receptors from the cell body along dendritic spines and finally undergo exocytosis onto the spine membrane.  Typical endosomes have diameters around 1-2 \textmu m whereas spine necks have diameters of several hundred nm, and the forces driving translocation are generated by families of molecular motors that move bidirectionally \cite{Park2013}.   The diversity of molecular constituents helps account for the rich set of behaviors observed experimentally in fluorescently labeled vesicles transiting into and out of the spine. These behaviors include positive and negative translocation velocities, and step-like transitions between these velocities  \cite{Da2015} suggestive of a discrete set of possible values. Additionally, the small aspect ratio of the long narrow dendritic spines causes the vesicles to deform as they squeeze through the neck, as can be seen in the timelapse images and kymograph in Fig.~\ref{figure1}(a) and (b).    Although much progress has been made on cataloguing the types of cytoskeletal molecules present \cite{kneussel2013myosin,Da2015}, such as molecular motors, the interplay of geometry, hydrodynamics and motor activity that governs how vesicles are forced through the spine neck is still poorly understood.

%Snapshots were taken at 1 second intervals for a total of 2 minutes and 10 seconds. We have tracked the vesicle using ImageJ and obtained timeseries of the trajectory and velocity in the $z$-direction. The data has been smoothed using a five-point moving average to remove high-frequency noise.   generously provided by the Kapitein lab at Utrecht University. Performing image analysis on a representative \emph{in vivo} recording of a fluorescently labeled vesicle transiting into and out of the spine (Fig. \ref{figure1} (c)), we observe both positive and negative velocities in the smoothed data and step-like transitions between velocities suggestive of a discrete set of possible values (Fig.~\ref{fig:spine_sim}(a)). 

 To begin analyzing this system, we consider a simple model of translocation, wherein an elastic vesicle of radius $R_p$ is driven by an applied force $F$ through a rigid constriction in a channel filled with a Newtonian fluid. We define the constriction by a single scale, e.g. a radially symmetric bump of radius $R_c$ at its narrowest point (Fig.~\ref{figure1}(c)). We define the confinement ratios $\pi_1$ by $\pi_1 := R_p/R_c$ and the dimensionless forcing $\pi_2$ by $\pi_2 := CF/(\pi R_p^3)$, where the compliance $C$ represents the elasticity of the vesicle, so that the two dimensionless quantities $\pi_1$ and $\pi_2$ govern the translocation dynamics. In this work, we restrict attention to confinement ratios $\pi_1 < 1$, since this case already involves significant computational challenges in resolving the lubrication layer and since nearly spherical vesicles allow for several simplifying theoretical assumptions. \edrs{In the Discussion section, we comment on why lessons learned from studying this case are likely to apply more generally to the case of extreme confinement ratios $\pi_1 \gg 1$ in which the vesicle must undergo large deformations simply to fit inside the channel.}
 
 %We first consider Sections \ref{sec:LB} and \ref{sec:lube} by assuming an external, prescribed force on the vesicle and neglect the velocity-dependent forcing of motor molecules (Fig.~\ref{figure1}(d)). In Section \ref{sec:compex}, we return to the particular example of endosome trafficking in dendritic spines by incorporating force generation by molecular motors and investigate the dynamics in this special case.

 \subsection{Lattice Boltzmann model} 
\label{sec:LB}

Before moving to a simplified mathematical description of this problem, it is useful to perform full-scale 3D simulations to get a sense of the elastohydrodynamical effects associated with the motion of a soft vesicle through a narrow constriction. %The lattice Boltzmann method is a natural choice for this investigation since it can be parallelized efficiently for use on computer clusters.
 We use the lattice Boltzmann approach as our fluid solver and  a finite element method to compute the elastic response of the vesicular membrane. These are coupled via an explicit immersed boundary method for the interaction of the fluid and the elastic membrane. The 3D lattice Boltzmann solver  considers the fluid as a cluster of pseudo-particles that move on a lattice under the action of external forces. The fluid is represented by a distribution function $\vec{f}_i$ that represents the probability of finding a pseudo-particle at position $\vec{r}$ with velocity in direction $\vec{e}_i$. The position and velocity spaces are both discretized on a lattice with spacing $\delta x$, and we use the so-called D3Q19 lattice, which has nineteen velocity directions \cite{benzi1992, qian1992}. The time evolution of the force distribution ${f}_i$ is governed by
\begin{equation}
f_i\left( \vec{r} + \vec{e}_i \delta t, t + \delta t \right) -f_i \left(\vec{r}, t \right) = \Omega_i,
\end{equation}
where $\delta t$ is the discrete time step and $\Omega_i$ is the collision rate between the fluid pseudo-particles, which is approximated by the Bhatnagar-Gross-Krook operator, $\Omega_i = -\delta t (f_i - f_i^{eq})/\tau_r$. Here $\tau_r$ is the relaxation time, related to the dynamic viscosity by $\mu = \rho c_s^2 \delta t (\tau_r/\delta t - 1/2)$, where $c_s = 1/\sqrt{3} \delta x/ \delta t$ is the lattice speed of sound and $\rho$ is the density of the fluid. One can convert between lattice units and SI units using these relations; for our calculations, we set the lattice constant, timestep, mass, and relaxation time to unity (see \cite{Janoschek2010} for more information on the conversion between physical and lattice units). 

The equilibrium distribution is given by the truncated Maxwell-Boltzmann distribution,
\begin{equation}
f_i^{eq} (\rho, \vec{u}) = \omega_i \rho \left[ 1 + \frac{\vec{c}_i \cdot \vec{u}}{c_s^2} + \frac{(\vec{c}_i \cdot \vec{u})^2}{2 c_s^4} - \frac{\vec{u} \cdot \vec{u}}{c_s^2}\right],
\end{equation}
where $\vec{u}$ is the velocity vector and the $\omega_i$ are weight factors that result from the velocity space discretization. Using this relation, we can calculate macroscopic hydrodynamic quantities such as the local pressure and velocity.  
 
 \subsection{Membrane model and geometry}
 
Deviations from the equilibrium shape of the vesicle increase its total elastic energy, which consists of in-plane shear and area dilation terms, and an out-of-plane bending term. The total energy is thus given by the sum of these contributions:
\begin{equation}
E_S + E_B =  \int \epsilon_s d A +  \frac{\kappa_B}{2} \int H^2 d A,
\end{equation}
where $\epsilon_s= \kappa_s \left( I_1^2 + 2 I_1 - 2 I_2\right)/12 + \kappa_{\alpha} I_2^2/12$ is the surface strain energy density, which depends on the surface elastic shear modulus $\kappa_s$ and the area dilation modulus $\kappa_{\alpha}$ as in \cite{Krugerthesis,kruger2014interplay}, while $\kappa_B$ is (the out-of-plane) bending modulus of the membrane and $H$ is the mean curvature. Here $(\lambda_1,\lambda_2)$ are the eigenvalues of the displacement gradient tensor $D$,  the invariants $I_1$ and $I_2$ are defined by $I_1 = \lambda^2_1 + \lambda^2_2-2$ and $I_2 = \lambda^2_1\lambda^2_2-1$, and we have used  the nonlinear strain energy density proposed by Skalak \cite{Skalak1973} for biological membranes, which is valid for both small and large strains. We assume that the vesicle is slightly stretchable, with a maximum of $5 \%$ change in total area. For a more in-depth overview of both the LB method and the membrane model, including its relation to the microscopic structure, the numerical evolution of the deformation gradient, and the resulting membrane forces, we refer the reader to \cite{Kruger2011, Krugerthesis,kruger2014interplay}. 
%and assume that the bending stiffness is relatively small; as the vesicle is resistant to shear, it behaves more like a polymeric capsule than a bilayer membrane. 
\begin{table}[h!]
  \centering
  \caption{\edrs{Lattice Boltzmann parameters.}}
  \label{tab:tableLB}
  \begin{tabular}{cccccc}
    \toprule
    Symbol & Definition & Value in lattice units & \multicolumn{2}{c}{Value in SI units}& \\
    \midrule
  
    $\delta t$ &Time step & 1 & 4.4 $\times 10^{-10}$ & s  \\
    $\delta x$ &Lattice constant & 1 & $2.3 \times 10^{-8}$ & \textmu m \\
    $\mu$ & Fluid dynamic viscosity & 1/6 & $2 \times 10^{-4}$ & $\text{Pa}\cdot\text{s}$ \\
    $\rho$ & Fluid mass density & 1 & $1 \times 10^{3}$ & kg/m$^{3}$ \\
    %$c_s$ & Speed of sound & $1/\sqrt{3}$ & 30  & m/s  \\
    $F_\text{ex}$ & Applied force & 1 -- 5 & (1.5 -- 7.5) $\times 10^{-9}$ & N \\
    $\kappa_{\alpha}$ & Area dilation modulus & 1 & 0.06 & N/m \\
    $\kappa_{s}$ & Shear modulus & 0.015 & $1 \times 10^{-3}$ & N/m\\
    $\kappa_{B}$ & Bending modulus & 0.018 & $5 \times 10^{-19}$ & $\text{J}$ \\
    $R_{c}$ & Channel radius & 30 -- 40 & ($6.9$ -- $9.2) \times 10^{-1} $ & \textmu m \\
    $R_p$ & Vesicle radius & 15 -- 32 & ($3.5$ -- $9.4) \times 10^{-1} $ & \textmu m \\
    $L$ & Domain length & 384 & 8.9  & \textmu m \\
    
    \bottomrule
  \end{tabular}
\end{table}

  \subsection{Simulation results}

To address the question of how the applied force and relative vesicle size affect translocation, we focus on a vesicle with fixed mechanical properties as given in Table \ref{tab:tableLB}. The parameters were chosen to assure numerical stability, but also to yield dimensionless values governing translocation that are comparable to typical values for bilayer vesicles. Note that the dimensional values of the moduli and applied force are higher than the ones of typical bilayer vesicles \cite{boal2012mechanics}. However, previous research \cite{Kusters2014} has shown that although the energetics of highly stretchable containers are very different than in the case of bending energy-dominated containers, the resulting phase diagrams are similar in terms of translocation dynamics, suggesting some form of universality between the two limits. In our simulations we assume that the fluids inside and outside the capsule are Newtonian and have the same material properties, i.e. the same viscosity and density. Table \ref{tab:tableLB} shows the parameters used in our lattice Boltzmann simulations, both in dimensionless and SI units. \edrs{Note that these values may be manipulated by varying the viscosity and the speed of sound of the medium; this is discussed in more detail in Narv{\'a}ez et al.~\cite{Narvaez2010}. The Reynolds number of these simulations is on the order of 0.1 -- 1. Moreover, since the viscosity-dominated regime of lubrication theory is applicable whenever $h/R_p \ll 1$ and $(h/R_p)^2 \text{Re} \ll 1$ \cite{Acheson1990}, and in our case $h/R_p \approx 0.1$ and $(h/R_p)^2 \text{Re} = 10^{-3}$ -- $10^{-2}$, we are indeed in the non-inertial lubrication regime.}

\begin{figure}
\centering
\includegraphics[width=0.7\textwidth]{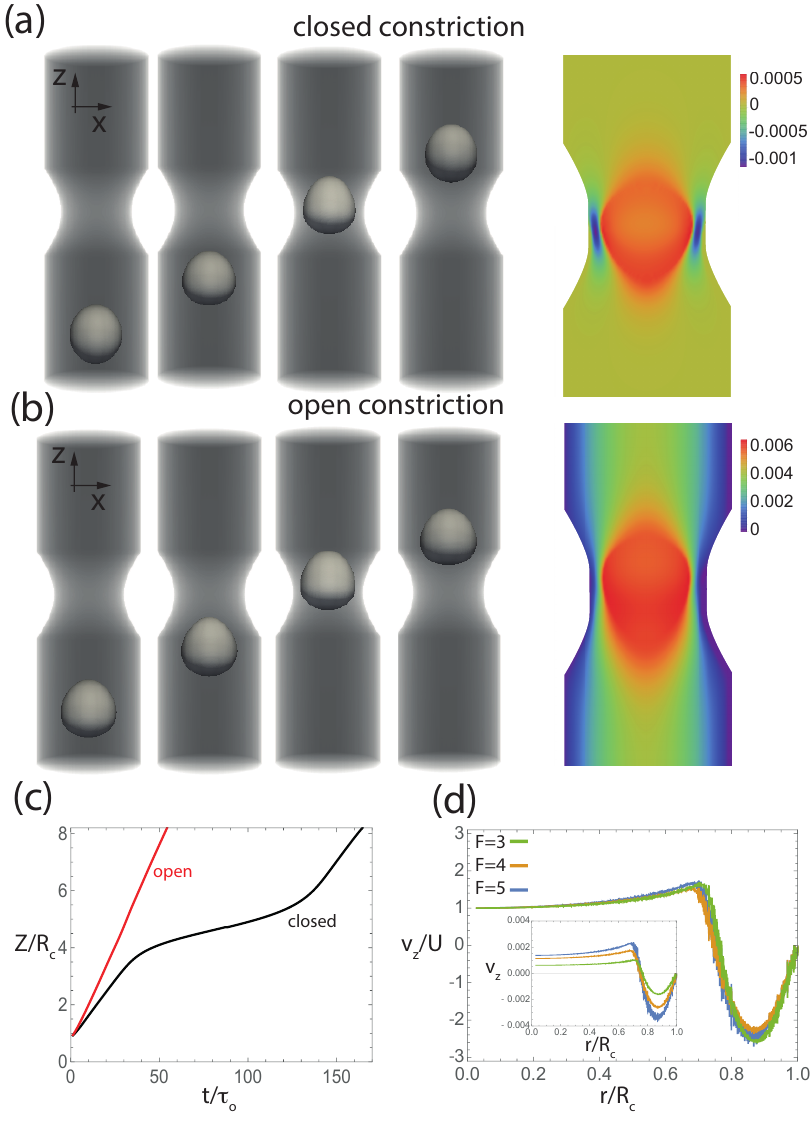}
\caption{(Color online) Translocation sequence and color-map of the $z$-velocity in the $x$-$z$ plane through a closed (a) and open (b) constriction for a vesicle with radius $\pi_1=0.75$ and applied dimensionless forcing $\pi_2=0.13$. While the shape of the vesicles during the translocation sequence is not markedly different, the associated velocities are. The color map in (a) of the $z$-velocity in the $x$-$z$ plane indicates the occurrence of a narrow fluid layer with reverse flow in the closed constriction that does not occur in the open constriction. (c) Time-evolution of the dimensionless position of the center of mass of the vesicle $Z/R_c$ passing through open (red) and closed (black) constrictions with otherwise identical parameters. Note that the velocity of the vesicle for the open constriction is only 5\% slower in the narrow part of the channel, whereas for the closed constriction it is 5 times slower.  (d) Fluid $z$-velocity $v_z$ in the closed constriction relative to the vesicle velocity $U$ as a function of $r/R_c$, the radial coordinate $r$ relative to the radius of the constriction, for applied forces $F=3, 4,$ and 5 corresponding to $\pi_2 = 0.16, 0.21,$ and 0.27 respectively. The data collapses in dimensionless coordinates, as evidenced by comparison to the inset, which shows dimensional velocities at different forcing strengths. \label{figure2}}
\end{figure} 

The surface of the vesicle is triangulated to allow for efficient calculations of the deformations, and the number of faces has been fixed to 3380, which is sufficient to capture the deformations studied (see the convergence study in Appendix \ref{app:LBsim}). We present our results in conventional lattice units ($\delta x=\delta t = 1$) in terms of the two dimensionless parameters mentioned above; the dimensionless confinement ratio $\pi_1 := R_p/R_c$ and the dimensionless force $\pi_2 := CF/(\pi R_p^3)$ where we choose area dilation as the dominant energetic contribution, hence $\pi_2= 5F/(\pi \kappa_{\alpha} R_p)$.

We first assume that there is a constant body force on the vesicle.  The total body force $F$ is implemented by applying a constant force uniformly to each node of the vesicle. To allow for the study of the lubrication layer between the vesicle and the channel wall, the fluid grid is made to be sufficiently fine, with size $196\times196\times384$, and we consider two different minimal neck radii with $R_c=30, 40$.  Figs.~\ref{figure2}(a) and (b) show a representative sequence of vesicle deformations and the associated velocity field during translocation for both closed and open constrictions with $\pi_1= 0.75$ and $\pi_2 = 0.13$. 

\subsection{Reverse flow due to a blind channel hinders translocation}

To showcase the difference between transportation through open and closed constrictions, we compare the time-evolution of the position of the center of mass of a vesicle in these two situations (Fig.~\ref{figure2}(c)). The dimensionless transit time through the constriction $\tau/\tau_0$ (where $\tau = R_p/U_\text{min}$ is measured relative to the transit time $\tau_0$ in free space) is up to an order of magnitude slower for the closed constriction than for the open constriction. As we will show further at the end of this section, the transit time $\tau$ sensitively depends on $\pi_1$, the vesicle radius relative to that of the constriction, and on the dimensionless applied force $\pi_2$. To explain this dramatic difference in translocation dynamics between open and closed constrictions, we now turn our attention to the boundary layer associated with the back-flow of fluid that occurs during translocation. 

To visualize the fluid flow in the narrow region between the channel wall and the vesicle,  in Figs.~\ref{figure2}(a) and (b) we plot the axial velocity (in the $z$-direction) of the fluid and vesicle for both the open and closed geometries. For the open constriction, the surrounding fluid is dragged along with the vesicle and no back-flow of fluid occurs. For the closed constriction, however, fluid incompressibility demands that a narrow layer with reverse flow emerges to allow fluid to escape the pocket as the vesicle enters the constriction. This reverse flow hinders the passage of the vesicle and increases the amount of deformation the vesicle must undergo in order to squeeze through the constriction. In Fig.~\ref{figure2}(d) we take a closer look at the velocity profile within the constriction and find that, by increasing the applied force, the amount of fluid that leaves the closed constriction per unit time (and hence the fluid velocity in the lubrication layer) increases as well. We note that rescaling the velocity of the surrounding fluid by that of the vesicle allows us to the collapse the velocity profiles onto a universal profile. 

We remark that the emergent lubrication layer is cylindrically symmetric, as one would expect from the presence of a lift force that drives the deformable vesicle from the surrounding walls \cite{cantat1999lift}. In Appendix \ref{app:LBsim} we also demonstrate this symmetry and measure the stability of the axisymmetric configuration by showing that the vesicle recovers its centered position in the constriction after small perturbations toward the wall.  

\subsection{Scaling of transit time}
\label{sec:LBscale}
\begin{figure}
\centering
\includegraphics[width=0.92\textwidth]{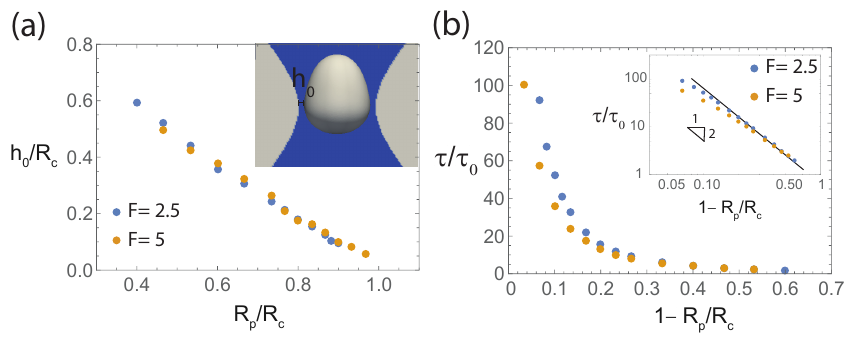}
\caption{(Color online) (a) The relative minimial gap size $h_0/R_c$ as a function of $\pi_1 = R_p/R_c$ for $F=2.5$ and 5, where $h_0$ is defined to be the minimal lubrication gap  during translocation (see inset). (b) The relative transit time $\tau/\tau_0$ as a function of $R_p/R_c$ for $F = 2.5, 5$. Plotting the same data on a log-log scale indicates that $\tau/\tau_0 \sim (h_0/R_c)^{-2}$ for gap sizes between $0.05< h_0/R_c < 0.5$ (inset), \edrs{though we cannot definitively establish a power law behavior from the full 3D simulations since the gap sizes range only over a single decade.} Here the vesicle stiffness is kept constant in all simulations and the corresponding dimensionless forcing is $\pi_2 = 0.1-0.5$. \label{figure3}}
\end{figure}

Our simulations show that the translocation of vesicles into closed constrictions is highly dependent on the driving force $F$ and the geometry of the vesicle and constriction via the radii $R_p$ and $R_c$, as illustrated in the inset Fig.~\ref{figure3}(a). In Fig.~\ref{figure3}(a) we measure the minimal lubrication gap $h_0$ between the vesicle and the channel as a function of $R_p$. For the case $R_p/R_c \ll 1$ this is a linearly decreasing function, but  for $R_p/R_c \approx 1$, the size of the lubrication layer $h_0/R_c$ no longer decreases linearly with $R_p/R_c$. Rather, it is set by a balance between the fluid pressure exerted on the membrane and the elastic properties of the vesicle. 

To quantify how this lubrication layer impacts the transit time $\tau$, we have measured $\tau$ for various values of the body force and confinement ratio $R_p/R_c$. As expected, increasing the radius of the vesicle lowers the minimal velocity in the constriction and hence increases $\tau$. If we rescale $\tau$ by the time $\tau_0$ it takes for a vesicle to move its own length (i.e.\ $\tau_0 = 6 \pi\mu R_p^2/F$ according to Stokes law), we find that all the curves collapse for moderate values of $R_p/R_c$ where $R_p/R_c <1$ (Fig.~\ref{figure3}(b)). For $1- R_p/R_c  \rightarrow 0$ we find that $\tau/\tau_0$ depends strongly on $R_p/R_c$, i.e. a slight increase in vesicle size drastically increases the passage time. \edrs{In the inset of Fig.~\ref{figure3}(b) we indicate that $\tau/\tau_0 \sim (h_0/R_c)^{-2}$ for gap sizes between $0.05 < h_0/R_c < 0.5$. Although we cannot definitively establish a power law behavior from the full 3D simulations since the gap sizes range over only a single decade, we note that this power of 2 is slightly smaller than the power of 2.5 for the rigid vesicle case that will be derived using the lubrication model in Section \ref{sec:lube_constf}.}
 
\edrs{We are prevented from studying this dependency for $R_p/R_c > 0.95$ since doing so would demand resolving very narrow fluid layers.} Note that in the regime $R_p/R_c >1$ in which undeformed vesicles do not fit within the constriction, it has been shown previously that for $R_p/R_c$ sufficiently large the vesicles get stuck in the constriction, i.e.\ the scaled transit time $\tau/\tau_0$ diverges \cite{Kusters2014}.

% We find $\tau/\tau_0 \sim (h_0/R_c)^{-2}$ for gap sizes between $0.05< h_0/R_c < 0.5$, \edrs{though we cannot definitively establish a power law behavior since the gap sizes range only over a single decade}
\section{Lubrication Model}
\label{sec:lube}
As outlined in the previous section, the transportation of vesicles through narrow, closed constrictions involves the emergence of a thin lubricating layer. Such layers are challenging to resolve in fixed-grid simulations such as the lattice Boltzmann method described above. However, we can take advantage of results from lubrication theory valid precisely in this limit, building on previous formulations for the motion of pellets through fluid-filled tubes  \cite{lighthill1968pressure,tozeren1978steady}.

Defining $h(z)$ to be the height of the vesicle above the channel wall as a function of the axial coordinate $z$, we assume axisymmetric flows and deformations, as justified by the results of 3D simulations presented in Section \ref{sec:LB}. The coordinate $z$ is defined so that $z \in (Z-R_p, Z+R_p)$, where $Z$ is the center of mass of the vesicle. Since $h \ll R_c$ in the case of interest, we may invoke the standard result from lubrication theory \cite{Acheson1990} that
\begin{equation}
 u(z) = \frac{1}{2\mu}\frac{\partial p}{\partial z} r(r-h)+\frac{U}{h}r,
\end{equation}
where $U=\ud Z/\ud t$ is the vesicle velocity in the $z$-direction. By incompressibility, the flux $Q$ through the gap must be equal through each cross section \cite{Acheson1990} so
\begin{equation}\label{eqn:q}
 Q=2\pi R_c\left(\frac{-h^3}{12\mu}\frac{\partial p}{\partial z}+\oh Uh\right)=\text{constant}.
\end{equation}
%In this section, we assume that $R_c \approx R_p$ and we refer to them interchangeably as $R$ unless specified.
Rewriting \eqref{eqn:q} in terms of $\partial p/\partial z$ and integrating,
\begin{equation}\label{eqn:p}
 \frac{p(z)-p_0}{6\mu}=U\int_{Z-R_p}^{z} \frac{1}{h^2(s)}\ud s-\frac{2Q}{2\pi R_c}\int_{Z-R_p}^{z} \frac{1}{h^3(s)}\ud s.
\end{equation}
Setting $z=Z+R_p$ in the above equation \eqref{eqn:p} determines the flow rate $Q$ in terms of the pressure drop $p(Z+R_p)-p_0=\Delta p$, which is a function of the applied force $F$ via $\Delta p = F/(\pi R_p^2)$, neglecting the viscous drag term of order $\mu U R (R/h)$ from the balance of forces since it is dominated by the pressure scale $\mu U/\rho (R/h)^2$ in the lubrication limit. (For the purpose of this scaling argument, $R \sim O(R_c) \sim O(R_p)$ as we are interested in the limit $R_c \approx R_p$.) This results in the equation
\begin{equation}\label{eqn:pL}
  Q = 2\pi R\left(U\int_{Z-R_p}^{Z+R_p} \frac{1}{h^2(s)}\ud s -\frac{F}{6\pi R^2\mu}\right)\Bigg/\left(2 \int_{Z-R_p}^{Z+R_p} \frac{1}{h^3(s)}\ud s\right).
\end{equation}
By conservation of mass, we know that the backflow $Q$ is balanced by the fluid dragged forward by the vesicle, i.e.\
\begin{equation}\label{eqn:u}
 Q = -\pi R_c^2 U,
\end{equation}
where we have assumed that the gap is sufficiently small that $R_p \approx R_c$. To close the systems of equations for $p(z)$, $h(z)$, $Q$, and $U$, we need a constitutive law relating the height to the pressure. We assume an approximately spherical vesicle and make the ansatz  that
\begin{equation}\label{eqn:h}
 h(z)=\widetilde{R}_c(z)-\sqrt{R_p^2-(z-Z)^2}+C (p(z)-p_0),
\end{equation}
where the channel radius $\widetilde{R}_c(z)$ is a function of position (related to the minimum channel radius via $R_c = \min_z \widetilde{R}_c(z)$, $C$ is the elastic compliance of the vesicle, and $p_0$ is the far-field pressure which we assume vanishes.  Note that \eqref{eqn:h} is a drastically simplified form of the vesicle elasticity since it does not allow for axial extension and neglects constraints such as conserved vesicle volume. A derivation of the compliance $C$ based on the vesicle material properties is given in Appendix \ref{app:comp}.

%In \eqref{eqn:h}, $\eta(x)$ is a mollifier function that satisfies $\eta(\pm R) = 0$, and $p_0$ is a constant that enforces the constraint of constant vesicle volume. Equations \eqref{eqn:p}--\eqref{eqn:h} fully specify the motion of the vesicle. The method for determining $p_0$ is described in Appendix \ref{app:height}, and the derivation of the compliance $C$ based on the vesicle material properties is contained in Appendix \ref{app:comp}.
  
Equations \eqref{eqn:q}-\eqref{eqn:h} give a system of four equations for the the unknown flux $Q$, and velocity $U$, the pressure $p(z)$, and the height $h(z)$, and fully describe the dynamics of the forced vesicle in the constriction. (The equivalent non-dimensionalized system is presented in Appendix \ref{app:nondim_lubem}.) Rather than solving \eqref{eqn:q}-\eqref{eqn:h}simultaneously, we start with the vesicle position $Z=Z(t)$ and initial guesses $p^0(z)$, $h^0(z)$, $Q^0$, and $U^0$ and use an iterative method as follows:
\begin{align}
 \frac{p^{i+1}(z)-p_0}{6\mu}&=U^i\int_{Z-R_p}^z \frac{1}{(h^i)^2(s)}\ud s-\frac{2Q^i}{2\pi R_c}\int_{Z-R_p}^z \frac{1}{(h^i)^3(s)}\ud s,\\
  h^{i+1}(z)&=\widetilde{R}_c(z)-\sqrt{R_p^2-(z-Z)^2}+Cp^i(z),\\
  Q^{i+1} &= 2\pi R_c\left(U^i\int_{Z-R_p}^{Z+R_p} \frac{1}{(h^i)^2(s)}\ud s -\frac{F}{6\pi R_p^2\mu}\right)\Bigg/\left(2 \int_{Z-R_p}^{Z+R_p} \frac{1}{(h^i)^3(s)}\ud s\right), \label{eqn:qit}\\
  U^{i+1} &= -Q^i/(\pi R_c^2), \label{eqn:uit}
\end{align}
iterating until the system converges to fixed points $p^\infty(z)$, $h^\infty(z)$, $Q^\infty$, and $U^\infty$. The vesicle position is then updated using $\ud Z/\ud t = U^\infty$ and the process is repeated using the new position $Z(t+\Delta t)$. The converged values $p^\infty(z)$, $h^\infty(z)$, $Q^\infty$, and $U^\infty$ at the previous step are used as initial guesses for the next iteration. This ensures that the iterates converge reliably to a solution provided that the timestep $\Delta t$ is sufficiently small. We have found that the rate of convergence can be accelerated using Steffensen's method \cite{burden2004}.

\begin{table}[h!]
  \centering
  \caption{Lubrication model parameters.}
  \label{tab:table1}
  \begin{tabular}{cccc}
    \toprule
    Symbol & Definition & Value & Units\\
    \midrule
    $\mu$ & Fluid viscosity & $1.2 \times 10^{-3}$ & $\text{Pa}\cdot\text{s}$\\
    $\rho$ & Fluid density & $1 \times 10^{3}$ & $\text{kg}/\text{m}^3$\\
    $F_\text{ex}$ & Applied force &$40$ -- $200$ & pN\\
    $C$ & Compliance & $5 \times 10^{-9}$ & $\text{m}/\text{Pa}$\\
    $R_{c}$ & Channel radius & $1.22$ -- $2.15$ & \textmu \text{m}\\
    $R_p$ & Vesicle radius & $0.96$ -- $1.5$ & \textmu \text{m}\\
    %$l_\text{wide}$ & Wide channel length & $9.6e^{-5}$ & \text{cm}\\
    $l_\text{narrow}$ & Narrow channel length & $2.5$ & \textmu \text{m}\\
    $l_\text{trans}$ & Transition length & $2.5$ & \textmu \text{m}\\
    %$h_\text{wide}$ & Largest channel radius & $9.6e^{-5}$ & \text{cm}\\
    %$h_\text{narrow}$ & Smallest channel radius & $9.6e^{-5}$ & \text{cm}\\
    \bottomrule
  \end{tabular}
\end{table}

\subsection{Scaling of transit time at constant applied force}
\label{sec:lube_constf}
\begin{figure}
\centering
%\includegraphics[scale=0.5]{lube_model_page_new4.png}
%\caption{(a) Schematic of the lubrication model: a nearly spherical vesicle passes close to the channel wall with a separation $h(x)$ and velocity $\mb{U}$ generated by an external force $\mb{F}$. (b) The trajectory obtained from the lubrication model is in reasonable agreement with the 3D immersed boundary simulation, where the parameters used in both simulations are contained in Table \ref{tab:table1}. (??difference??) (c) The compliance $C$ is fit to data from the 3D simulation and results in a consistent height profile. (d) The lubrication model was used to investigate the scaling of passage time with the vesicle size through the minimum clearance $h_0$. The scaling argument form Section \ref{sec:lube_constf} shows that $\tau \sim  h_0^{2.5}$ in the inelastic case with $C=0$. The lubrication model recovers this limit in the inelastic case; in the elastic case we find a weaker dependance on the height, as expected, with $\tau \sim  h_0^{2.05}$.
\includegraphics[scale=1.7]{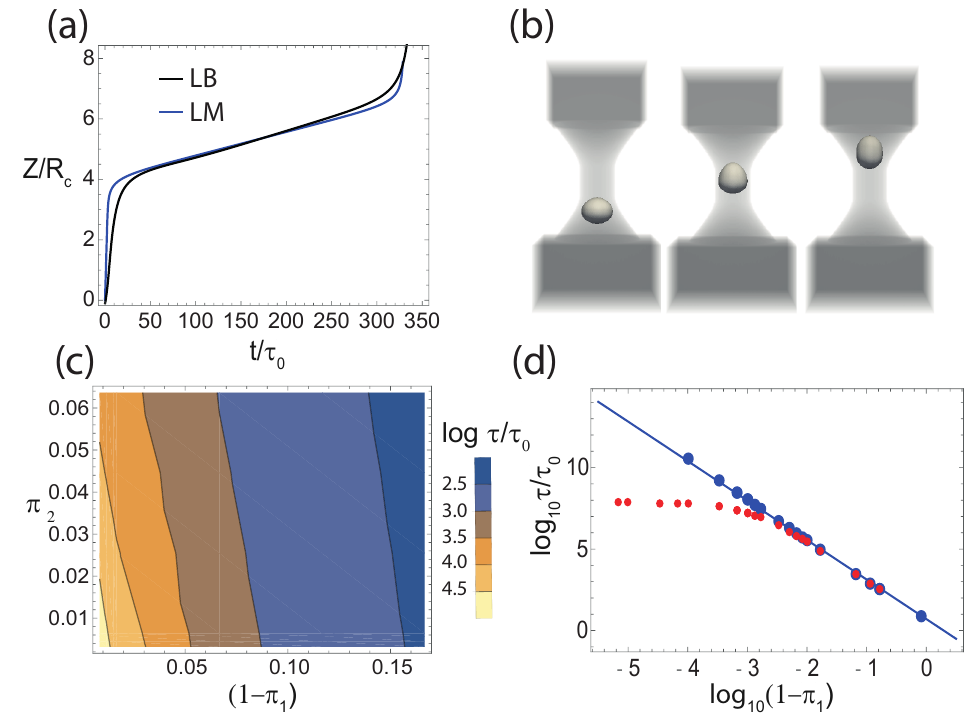}
\caption{(Color online) (a) The trajectory obtained from the lubrication model (LM) is in good agreement with the 3D lattice Boltzmann simulation (LB) with parameters from Tables \ref{tab:tableLB} and \ref{tab:table1}  \edrs{such that $\pi_1 = 0.83$ and $\pi_2 = 0.090$}. The discrepancy in the wide part of the channel is not surprising, since the assumptions of the lubrication model break down when $h \sim R_p$. (b) Snapshots during translocation from the 3D lattice Boltzmann simulations. (c) The lubrication model was used to investigate the scaling of the non-dimensional transit time $\tau/\tau_0$ with the relative vesicle size $1-\pi_1 \approx h_0/R_c$ and the normalized force $\pi_2$. (d) Transit time versus $1-\pi_1$ on a log scale in the limit $\pi_1 \to 1$. According to the scaling argument in Section \ref{sec:lube_constf}, $\tau \sim  h_0^{-5/2}$ \edrs{in the inelastic regime $\pi_2/2 \ll 1-\pi_1$} (blue line). The lubrication model recovers this limit in the inelastic case (blue symbols). In the elastic case with $\pi_2 = 1.8 \times 10^{-3}$  (red symbols), we find a weaker dependence on the height and a plateau with minimum height $h_0/R_p \approx \pi_2/2$ \edrs{at the transition $\log_{10}(1-\pi_1) = \log_{10}(\pi_2/2) \approx -3.0$.}}
\label{figure_lubemodel} 
\end{figure}

As shown in Fig.~\ref{figure_lubemodel}(a), there is good agreement between the trajectories computed by the lubrication model (LM) and 3D lattice Boltzmann (LB) simulations. Here, we have used the channel geometry illustrated in Fig.~\ref{figure_lubemodel}(b). Because of its analytical simplicity in comparison to directly solving the full 3D fluid-structure interaction problem, the lubrication model allows for a thorough exploration of the phase space associated with $\pi_1 = R_p/R_c$ and $\pi_2 = CF/(\pi R_p^3)$ (Fig.~\ref{figure_lubemodel}(c)). (See Table \ref{tab:table1} for the parameters used in the lubrication model.) For an elastic vesicle with compliance $C = 5 \times 10^{-9}\, \text{m}/\text{Pa}$, simulation of the lubrication model reveals a plateau in the transit times as $\pi_1 \to 1$ (Fig.~\ref{figure_lubemodel}(d)). The reason for this plateau is that, when $\pi_1 \sim 1$, the deflection term $C p(z)$ dominates in the equation \eqref{eqn:h} for the height. It is precisely $\pi_2$ that controls the magnitude of this deflection term: we find the dimensionless minimum height $h_0/R_c \approx \pi_2/2$.

\edrs{In the case $\pi_2/2 \ll 1-\pi_1$ in which elastic deformations are negligible}, the absence of deformations yields much narrower lubrication layers. We find a scaling relation $\tau/\tau_0 \sim  (h_0/R_c)^{-5/2}$ in the small $h_0$ limit. (Recall that $\tau_0$ is the time required for the forced vesicle to move a distance equal to its own length in free space and that $h_0$ is the minimal gap spacing, which in the inelastic case is related to $\pi_1$ by $h_0/R_c \approx 1-\pi_1$.) To understand this scaling, we start from \eqref{eqn:pL} and \eqref{eqn:u},  which yield 
\begin{equation}
 \frac{F}{6\pi\mu R_p^2}=U\left(\int_{Z-R_p}^{Z+R_p} \frac{1}{h^2(s)}\ud s+R_c\int_{Z-R_p}^{Z+R_p} \frac{1}{h^3(s)}\ud s\right).
\end{equation}
To estimate the integral $\int_{Z-R_p}^{Z+R_p} 1/h^n(z)\ud s$ for arbitrary $n$, we use the formula \eqref{eqn:h} for the height function, neglecting the elastic term, so that
\begin{align}
\begin{split}
 \int_{Z-R_p}^{Z+R_p} \frac{1}{h^n(s)} \ud s &=\int_{-R_p}^{R_p} \frac{1}{(R_c-R_p\sqrt{1-(s/R_p)^2})^n} \ud s\\
&\approx \int_{-R_p}^{R_p} \frac{1}{(R_c-R_p(1-(s/R_p)^2/2))^n} \ud s\\
&\approx\frac{1}{h_0^n} \int_{-R_p}^{R_p} \frac{1}{(1+(R_p/2h_0)(s/ R_p)^2)^n} \ud s\\
&\approx \frac{R^{1/2}}{h_0^{n-1/2}} \int_{-R_p/\sqrt{h_0 R_p}}^{R_p/\sqrt{h_0 R_p}} \frac{1}{(1+y^2/2)^n} \ud y
\approx \frac{R^{1/2}}{h_0^{n-1/2}},
\end{split}
\end{align}
where making the approximation $\sqrt{1+s^2} \approx 1+s^2/2$ is justified since the integrand becomes singular near $s = 0$ as $h_0 \to 0$. It follows that the timescale $\tau \sim R/U \sim \mu R^{7/2}(h_0^{-3/2}+Rh_0^{-5/2})/F \sim (\mu R^{9/2}/F) h_0^{-5/2}$, since when $h_0 \ll 1$, $h_0^{-3/2} \ll h_0^{-5/2}$. Since $R_c$ is constant and $\tau_0$ is constant in the limit $R_p \to R_c$, we have $\tau/\tau_0 \sim (h_0/R_c)^{-5/2}$.

\edrs{Neglecting the elastic term is appropriate since for $s \approx 0$ and $R_c \approx R_p =: R$,
\begin{align}
\begin{split}
\frac{h(s)}{R_c} &\approx \frac{R_c-R_p\left(1-s^2/(2 R_p^2)\right)+Cp(s)}{R_c} \\
&\approx 1-\pi_1 +\frac{s^2}{2 R^2}+\pi_2/2\\
&\approx 1-\pi_1 +\frac{s^2}{2 R^2},
\end{split}
\end{align}
where we have used $p(s) \approx F/(2 \pi R^2)$ at $s \approx 0$ in the second line and the assumption $\pi_2/2 \ll 1-\pi_1$ in the final line.}

\edrstwo{In contrast to the scaling of the transit time with the minimal gap size $\tau \sim h_0^{-5/2}$ derived in the inelastic case, the scaling observed in the lattice Boltzmann simulations of Section \ref{sec:LBscale} is $\tau \sim h_0^{-2}$. This is because those simulations take place in a regime that is neither inelastic nor dominated by membrane compliance, but rather in a transition regime, corresponding to the bend in the graph of Fig.~\ref{figure_lubemodel}(d), that gives rise to an apparent $h_0^{-2}$ scaling over a narrow range of $h_0$.}

%The viscous stress is of order $\mu U/h$, so that the net drag is of order $\mu U R^2/h$. Placing an external force $F$ in balance with the drag yields an estimate for the timescale $\tau$ of passage according to $\tau \sim R/U \sim \mu R^3/(\h F)$, so that the time to passage scales as the inverse of the height.

%pressure $P$ through the gap must be balanced by viscous damping, which yields the force balance $P/\sqrt{Rh} \sim \mu U/h^2$, where we have used that the vesicle is in near contact with the channel through a contact zone of order $\sqrt{Rh}$. Given a clearance $h_0$ in the pressure-free case, by the definition of the compliance $h \sim h_0+P/C$, so $P\sim\mu U R^{1/2}/(h_0+P/C)^{3/2} \sim \mu U R^{1/2}/(h_0^{3/2}(1+P/(h_0 C))^{3/2} \sim \mu (U R^{1/2}/h_0^{3/2}) (1/(h_0 C)) (\mu U R^{1/2}/h^{3/2})$. 

\subsection{Active translocation by molecular motors}
\label{sec:lube_motor}

Having seen how the simplicity of the lubrication model allows us to understand the pressure-driven passage of deformable vesicles through narrow intracellular channels relatively easily, we now turn to complement our analysis by accounting for endosomal transport driven by the dynamics of molecular motors that have their own kinetics of binding and unbinding associated with force production.  

%We are able to recapitulate these behaviors by using a two-species molecular motor model, in which the two species apply forces in opposite directions. Interestingly, including these two species at equal concentrations leads not to zero net force, but rather a bistable dynamics in which vesicles can be trafficked in either direction. This is an example of  Moreover, ...something about corking...

Although the molecular motor dynein walking along microtubules plays an important role in transporting vesicles through dendritic branches, there is strong experimental evidence that the actin-myosin cytoskeleton dominates transport into the spines \cite{Da2015}. Therefore, to model motor activity, we use a force-velocity relation for myosin binding to actin filaments in muscle based on the classic work of Huxley \cite{huxley1957muscle} and more recent work by Lacker and Peskin \cite{lacker_peskin,hopp_pesk}. Although the details of molecular motor function undoubtedly differ between bidirectional intracellular transportation and muscle, the key model assumptions do not depend sensitively on the details of the individual motors, in that we assume (i) there are two independent motor species that are identical aside from the direction in which they push, and (ii) motors are fixed within the cortex and do not diffuse or travel with the vesicle, as allowed in some other models such as \cite{yochelis2015self}. We also refer the interested reader to the related descriptions of molecular motor assemblies in \cite{julicher1997,campas2006collective}. Defining $\phi(z)$ to be the population density of motors attached to a vesicle and $\theta_{[a,b)}$ as the fraction of attached motors with displacements satisfying $a \le z < b$, it follows that
\begin{equation}
 \theta_{[a,b)} = \int_a^b \phi(z)\ud z.
\end{equation}
The total fraction $\theta$ of attached motors is therefore
\begin{equation}
 \theta = \int_{-\infty}^\infty \phi(z)\ud z,
\end{equation}
and the total force exerted on the vesicle is
\begin{equation}
\label{eqn:motorforce}
 F = n_0\int_{-\infty}^\infty f(z)\phi(z) \ud z,
\end{equation}
where $n_0$ is the total number of motors and $f(z)$ is the functional form of the force exerted by an individual motor, which will be specified later on. Assume rate constants $\alpha$ and $\beta$ of motor attachment and detachment, respectively, and that connections are always formed at a displacement $z=A$, so that the motors tend to push the vesicle \emph{in the $-z$ direction} with velocity $U < 0$. (We will later discuss the case of motion in the $+z$ direction.) Now, considering only those attachments with displacement $z_0 < z < A$ for fixed $z_0$, at steady-state it must follow that
\begin{equation}
\label{eqn:motorpde_short}
 \alpha(1-\theta)=\beta\int_{z_0}^A \phi(z)\ud z - U\phi(z_0),
\end{equation}
representing a balance between formation of new attachments at a rate $\alpha(1-\theta)$, detachment at a rate $\beta\int_{z_0}^A \phi(z)\ud z$, and motion by advective flux at a rate $-U\phi(z_0)$. From \eqref{eqn:motorpde_short}, one can derive the equation
\begin{equation}
\label{eqn:motorpde_short_pde}
 U \ud \phi/\ud z = -\beta \phi
\end{equation}
 as well as the boundary condition $\alpha(1-\theta) = -U\phi(A)$, from which it follows that
\begin{equation}
 \phi(z) = \frac{\alpha\beta}{-U(\alpha+\beta)}\exp\left(-\beta(z-A)/U\right).
\end{equation}
Using the form  $f(z) = -p_1\left(\exp\left(\gamma z\right)-1\right)$ for the force exerted by an individual motor, we substitute into \eqref{eqn:motorforce} the above result and carry out the integration in \eqref{eqn:motorforce} to obtain the force-velocity relation for the ensemble of motors 
  \begin{equation}
   F = -\frac{\alpha n_0 p_1}{\alpha+\beta}\frac{\left(e^{\gamma A}-1\right)+\left(\gamma U/\beta\right)}{1-\left(\gamma U/\beta\right)} \label{force_vel1},
  \end{equation}
which can be rewritten in the familiar Hill-like form \cite{hopp_pesk,hill1938heat} as
\begin{equation}
 U = \frac{b(F_0-F)}{F+a},
\end{equation}
where
\begin{align}
 a &= -\alpha n_0 p_1/(\alpha+\beta),\\
 b &= 2\beta/\gamma, \\
 F_0 &= -\frac{\alpha n_0 p_1}{\alpha+\beta}\left(e^{\gamma A}-1\right).
\end{align}

%For the motors forming connections at displacement $z=A$, motion in the $-z$ direction corresponds to \emph{shortening} in the muscle context, since each myosin motor tends to decrease its displacement over time.

\begin{figure}
\centering
\includegraphics[scale=0.75]{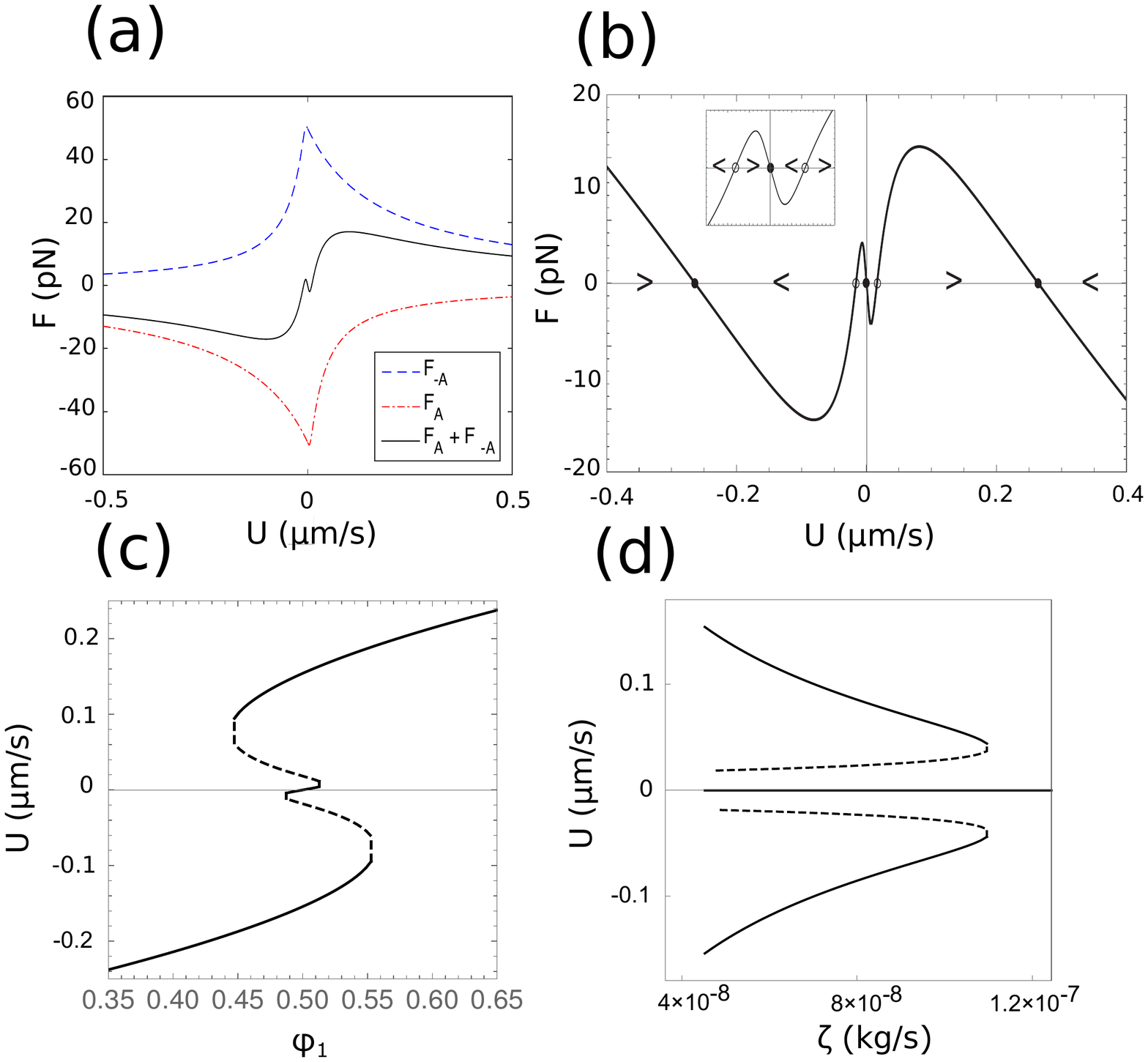}
\caption{(Color online) \edrs{Bifurcation analysis of competing motor species pushing a rigid vesicle through a constriction into a blind end.} (a) Force-velocity curves for a single motor species forming connections at $z=-A$ (blue, dashed) and a single motor species forming connections at $z=A$ (red, dashed/dotted) along with the sum of the two force-velocity curves (black, solid). We have set the value of the dimensionless parameter $\pi_5 := \gamma(B-A)$ to be $\pi_5 = 0.1$. (b) Steady-state solutions obtained by plotting the difference between the motor force-velocity curve and the viscous drag $F = \zeta U$ for friction coefficient $\zeta \approx 4.5 \times 10^{-8}$ kg/s. For reference, the friction coefficient in free space predicted by Stokes' law is $\zeta \approx 2.7 \times 10^{-8}$ kg/s. Arrows are drawn to illustrate the effect of small perturbations from steady states; e.g. perturbations from the leftmost steady state result in acceleration back toward the leftmost steady state, hence it is stable (solid circle) as opposed to the two unstable steady states (open circles) for which small perturbations give rise to divergent trajectories. The inset shows a close-up of the three steady states near the origin. (c) Bifurcation diagram showing the steady-state velocities as a function of the fraction of motors $\phi_1$ pushing up the channel. Solid lines are used to denote stable steady states while dashed lines are used for unstable steady states. (d) Bifurcation diagram showing the dependence of the steady-state velocities on the friction coefficient $\zeta$, under the assumption of equal fractions of motors pushing in each direction ($\phi_1=\phi_2=0.5$).}
\label{fig:figure_mixed}
\end{figure}

The case of motion in the $+z$ direction is posed as an exercise in \cite{hopp_pesk}, and we work it out here in detail. For $U > 0$, the equation corresponding to \eqref{eqn:motorpde_short} is
\begin{equation}
\label{eqn:motorpde_length}
 \alpha(1-\theta)=\beta\int_{A}^{z_0} \phi(z)\ud z + U\phi(z_0).
\end{equation}
If we assume that the rate $\beta$ of detachment is constant for all $z > A$, this results in the population density
\begin{equation}
 \phi(z) = \frac{\alpha\beta}{U(\alpha+\beta)}\exp\left(-\beta(z-A)/U\right).
\end{equation}
Using this population density and performing the integration in \eqref{eqn:motorforce} results in the force
  \begin{equation}
   \label{eqn:force_vel}
   F = -\frac{\alpha n_0 p_1}{\alpha+\beta}\frac{\left(e^{\gamma A}-1\right)+\left(\gamma U/\beta\right)}{1-\left(\gamma U/\beta\right)},
  \end{equation}
which happens to be identical to \eqref{force_vel1}, so that there is a single smooth function representing motion in both the $+z$ and $-z$ directions. However, as motors cannot be stretched to arbitrarily large displacement without breaking, we introduce a maximum displacement $B > A$ at which motors must detach. The fraction of attached motors then becomes
\begin{equation}
 \theta = \int_A^B \phi(z)\ud z = \frac{U}{\beta}\phi(A)\left(\exp\left(\beta(B-A)/U\right)-1\right),
\end{equation}
and solving for the population density from \eqref{eqn:motorpde_short_pde} yields
\begin{equation}
  \phi(z) = \frac{\alpha\beta c}{-U(\alpha+\beta c)}\exp\left(-\beta(z-A)/U\right),
\end{equation}
where the normalization constant $c = 1-\exp\left(\beta(B-A)/U\right)$. Performing the integral
\begin{equation}
\label{eqn:motorforce_length}
 F = n_0\int_{A}^B f(z)\phi(z)
\end{equation}
with $f(z)$ as before results in the force-velocity curve
  \begin{equation}
   F_A = -\frac{\alpha n_0 p_1 c}{\alpha+\beta c}\frac{\left(e^{\gamma A}\left(1-e^{-\beta(B-A)/U}e^{\gamma (B-A)}\right)-1\right)+\left(\gamma U/\beta\right)+ e^{-\beta (B-A)/U}\left(1-\gamma U/\beta\right)}{1-\left(\gamma U/\beta\right)}.
  \end{equation}
Note that \eqref{eqn:force_vel} is recovered in the limit $B\to\infty$. When $B$ is finite, the force-velocity curves for motion in the $+z$ and $-z$ directions are no longer described by a single equation, but it can be shown that the resulting piecewise formula is smooth at the transition $U=0$ \cite{hopp_pesk}. All together then, for motors forming connections at $z=A$ and attempting to move the vesicle in the minus $z$ direction, 
\begin{equation}
F_A = 
\begin{dcases}
      -\frac{\alpha n_0 p_1}{\alpha+\beta}\frac{\left(e^{\gamma A}-1\right)+\left(\gamma U/\beta\right)}{1-\left(\gamma U/\beta\right)},  & U < 0\\
      -\frac{\alpha n_0 p_1 c}{\alpha+\beta c}\frac{\left(e^{\gamma A}\left(1-e^{-\beta(B-A)/U}e^{\gamma (B-A)}\right)-1\right)+\left(\gamma U/\beta\right)+ e^{-\beta (B-A)/U}\left(1-\gamma U/\beta\right)}{1-\left(\gamma U/\beta\right)}, & U \ge 0,
\end{dcases}
\end{equation}
where we have assumed that motors connect at a fixed displacement $z= A$. All motors attach at a fixed rate $\alpha$, detach at fixed rate $\beta$, and are stretched or contracted according to the vesicle velocity. All motors detach upon extension to $z=B$. See Fig.~\ref{fig:figure_mixed}(b) for the form of the resulting force velocity curve using a representative value of the nondimensional quantity $\gamma (B-A)$.

We have introduced the notation $F_A$ to denote the force generated by the motor species that forms connections at $z=A$ considered thus far. A competing species that forms connections at $z=-A$ and disconnects at $z=-B$ behaves in the opposite way, with motion in the $+z$ direction corresponding to shortening, and motion in the $-z$ direction corresponding to lengthening. The equations in this case are
\begin{equation}
 F_{-A} = 
\begin{dcases}
  \frac{\alpha n_0 p_1 c}{\alpha+\beta c}\frac{\left(e^{\gamma A}\left(1-e^{\beta(B-A)/U}e^{\gamma (B-A)}\right)-1\right)-\left(\gamma U/\beta\right)+ e^{\beta (B-A)/U}\left(1+\gamma U/\beta\right)}{1+\left(\gamma U/\beta\right)}, & U < 0\\
      \frac{\alpha n_0 p_1}{\alpha+\beta}\frac{\left(e^{\gamma A}-1\right)-\left(\gamma U/\beta\right)}{1+\left(\gamma U/\beta\right)},  & U \ge 0.
\end{dcases}
\end{equation}
We assume that the two competing species of motors are coupled only through the resulting velocity of motion. This assumption allows us to compute the net force on a vesicle simply by adding the forces from each species. Define $\phi_1$ to be the fraction of motors that form connections at $z=-A$ out of the $n_0$ total motors, so that the motor species that forms connections at $z=A$ makes up the fraction $\phi_2 = 1-\phi_1$ of the total. This results in a force $F = \phi_1F_{-A}+\phi_2F_{A}$, with the resultant force-velocity curve shown in Fig.~\ref{fig:figure_mixed}(c).

Because of the kinetics of attachment and detachment, including two opposing species of molecular motors at equal concentrations leads to multistable dynamics in which vesicles can be trafficked in either direction or remain stationary. For example, setting $\phi_1=\phi_2=0.5$ results in a multistable behavior in which the motors can generate forward, reverse, or net zero motion (Fig.~\ref{fig:figure_mixed}). The steady-state velocities are precisely those at which the force production curve intersects the viscous drag curve. \edrs{To simplify the bifurcation analysis, we approximate the viscous drag curve by a line through the origin.} The slope of the resulting viscous drag line depends on the degree of confinement; in free space, this is the familiar Stokes law $F = 6\pi\mu R_p U$. The existence of multiple steady-states is illustrated in Fig.~\ref{fig:figure_mixed}(b) by plotting the difference between the motor force-velocity curve and the viscous drag $F = \zeta U$ for a representative friction coefficient $\zeta$. Visual inspection reveals the existence of 5 steady states, 3 of which are stable. The stability of the rightmost steady state can be assessed in terms of the sign of the derivative at steady-state. For negative derivatives, if the velocity is increased slightly, the drag force exceeds the motor force and the vesicle slows down; if the velocity is decreased slightly, the motor force exceeds the drag and the vesicle speeds up. Therefore, the rightmost steady-state is stable. Likewise, positive derivatives indicate unstable steady-states. 

As the fraction $\phi_1$ is increased or decreased, there is a sequence of bifurcations at which the multiple fixed points coalesce onto the single stable solution observed in the case of having only one motor species (Fig.~\ref{fig:figure_mixed}(c)). For small $\phi_1$, there is a single stable velocity and it is in the $-z$ direction. As the fraction increases, there is a saddle-node bifurcation and a stable velocity in the $+z$ direction emerges in addition to the stable negative solution. When $\phi_1 \sim 0.5$, a second saddle-node bifurcation gives rise to another stable solution that crosses from small negative to small positive velocities. In Fig.~\ref{fig:figure_mixed}(d) we plot a complementary bifurcation diagram showing the dependence of the steady-state velocities on the height-dependent friction coefficient $\zeta$, under the assumption of equal numbers of both species of motors. For small values of $\zeta$, there are five steady-state velocities, of which three are stable. As the drag increases, two saddle-node bifurcations annihilate four steady states, leaving only the stationary solution. 

\edrs{Note that the approximation of linear viscous drag is used only to study the qualitative behavior; in our actual simulations, we always solve for the fluid drag that emerges from the coupling between fluid, elastic, and motor dynamics.}

Multistability in mixtures of motors of opposing polarities has been previously observed in both experimental and theoretical studies \cite{guerin2010}. Gilboa et al.~\cite{gilboa2009} performed motility assays on an experimental actomyosin system with actin tracks of alternating polarities and observed bidirectional motion. Muller et al.~\cite{muller2008} studied theoretically the effect of antagonistic motor species exerting forces in opposite directions, which they called tug-of-war, and observed multistability upon simulating a master equation for the distribution of bound states. Hexner and Kafri \cite{hexner2009} also considered a similar tug-of-war scenario and analyzed the possible phase space, including bifurcation diagrams consistent with those of Fig.~\ref{fig:figure_mixed}. More recent work includes the effect of thermal fluctuations on switching between states \cite{miles2016bidirectionality} and tug-of-war in the context of searches for a random target \cite{newby2010random}. The present study builds on this previous work by investigating quantitatively the subtle interplay between the multistable dynamics of mixed motor systems and the effects of hydrodynamics associated with the confined geometry in which the cargo moves. Our results demonstrate that the steady state velocities depend not only on the fraction of motor species present, but also on the effective Stokes law. Since the viscous drag depends on the geometry of the constriction through which the cargo passes, this sets the force that must be generated by the aggregate of mixed-polarity motors. As the channel height varies, the velocity can jump abruptly from one branch of solutions to another, as we now demonstrate in the next section.

\section{Vesicle trafficking into dendritic spines}
\label{sec:compex}
As mentioned in the introduction, the dendritic spine is an example of a heterogeneous environment in which various biopolymers  interact with different species of molecular motors that traverse filaments in opposite directions and species that are able to reverse their direction of motion. This diversity helps explain the richness of behaviors seen in vesicle transit through spines, including (i) one-directional transit through the spine (ii) bidirectional transit up to the spine neck and back, and (iii) corking in the spine neck.

\begin{figure}
\centering
\includegraphics[clip,trim={0 2in 0 0},scale=0.7]{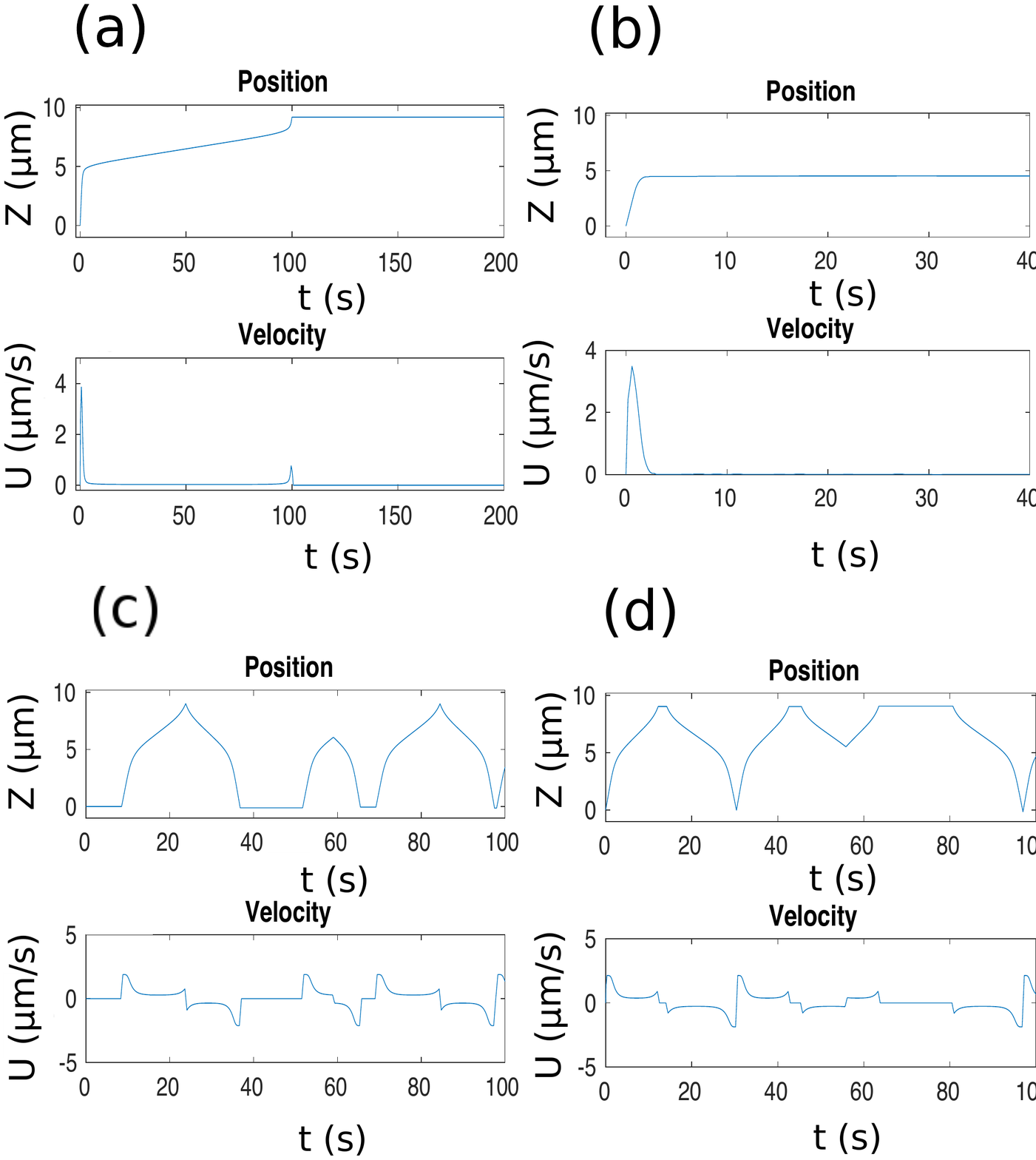}
\caption{Vesicle trajectory and velocity subject to different values of fractions $\phi_1$ of upward-directed motors. (a) The vesicle travels in a processive fashion to the tip of the spine with $\phi_1=0.57$, $R_p = 0.96$ \textmu m and $R_c = 1.22$ \textmu m, and $F_0 = 50$ pN. (b) The vesicle becomes trapped in the spine neck as it encounters a bifurcation in the steady-state solutions when $\phi_1 = 0.5$ and with all other parameters unchanged. (c) Demonstration of multistability in the lubrication model; both positive and negative velocities occur as in experiment \cite{Da2015} when using $\phi_1=0.46$, $R_p = 1.5$ \textmu m, $R_c = 2.15$ \textmu m, $F_0 = 200$ pN, and $\pi_5 = 0.02$. (d) Multistability is obtained using the opposite motor polarity with $\phi_1=0.54$, $R_p = 1.5$ \textmu m, $R_c = 2.15$ \textmu m, $F_0 = 200$ pN, and $\pi_5 = 0.02$.}
\label{fig:spine_sim}
\end{figure}

To model the effect of a crowded intracellular environment, we use an elevated fluid viscosity $\mu = 0.12\, \text{Pa}\cdot\text{s}$, which is 100x the viscosity of water. We also include a proxy for thermal fluctuations in the following manner: instead of taking the converged velocity $U^\infty_j$ at timestep $j$ as the initial guess for fixed point iteration at the next timestep  $U^0_{j+1}$, we add noise by setting $ U^0_{j+1} = U^\infty_j + \eta \xi$, where $\xi$ is a random number drawn from a zero-mean unit-variance normal distribution. This increases the likelihood that the solutions of fixed point iteration at subsequent time steps will lie on different solution branches. 

Upon including the effects of competing molecular motor species our model is able to recapitulate at a qualitative level many of the phenomena observed in vesicular trafficking into dendritic spines. \edrs{In these simulations $(\pi_2/2)/(1-\pi_1)$ ranges from $1/6$ -- $1/4$, so that according to the criterion of Section \ref{sec:lube_constf} vesicle elasticity plays only a moderate role and the qualitative analysis of Section \ref{sec:lube_motor} is applicable.} Our simulations lead to bidirectional motion and transit times on the order of $10$ -- $100$ seconds, in line with the experimental observations \cite{Da2015}. When the fraction $\phi_1=0.57$ of upward-directed motors is used, $R_p = 0.96$ \textmu m and $R_c = 1.22$ \textmu m, with a total motor force of 50 pN and noise magnitude $\eta = 1.35 \times 10^{-1}$ \textmu m/s, the vesicle moves in a processive fashion up the spine (Fig.~\ref{fig:spine_sim}(a)), consistent with the single steady solution in the limit $\phi_1 \to 1$. Given myosin's stall force of around 3 pN \cite{mehta1999myosin}, the motor forces of 50 -- 200 pN used here correspond to having a few dozen motor molecules. When equal fractions $\phi_1=\phi_2=0.5$ of upward and downward-directed motors are used with the parameters otherwise fixed, the vesicle becomes trapped in the spine neck (Fig.~\ref{fig:spine_sim}(b)). This is reminiscent of experimental observations of vesicles trapped in spines and hypotheses that such corking of vesicles in spine necks may serve as a chemical insulator \cite{Park2006,Wang2008}, as has been studied theoretically in terms of anomalous diffusion along spiny dendrites (see \cite{bressloff2013stochastic} and references therein).
When the fraction $\phi_1=0.46$ of upward-directed motors is used, either positive or negative velocities can occur at the same position in the channel depending on the random initial guess. In one representative simulation, the vesicle makes it through the spine neck on its first attempt but not on its second attempt, during which it turns around and must try again (Fig.~\ref{fig:spine_sim}(c)). This last simulation is done using a different set of parameters, with the minimum channel radius $R_c = 2.15$ \textmu m, vesicle radius $R_p = 1.5$ \textmu m, forcing $F_0 = 200$ pN, noise magnitude $\eta = 1.35 \times 10^{-2}$ \textmu m/s, and $\pi_5 = 0.02$. We find analogous behavior using equal but opposite fractions with $\phi_1=0.54$, but with the vesicle spending most of its time sequestered in the spine (Fig.~\ref{fig:spine_sim}(d)).

Our simulations are subject to the boundary conditions that only non-negative velocities are allowed at the base of the channel, and only non-positive velocities are allowed at the end of the channel. Therefore the vesicle waits at the base until it acquires a large enough random kick and begins to move in the $+z$ direction into the spine. Similarly, the vesicle stops at the end of the spine and waits until it receives a sufficiently large random kick to be brought in the opposite direction by the same assembly of motors. %Since the comparison to experiment here is done on a qualitative rather than a quantitative level, we have used the faster timescale in the lubrication model for ease of comparison to the results of our 3D simulations, which are limited to short times because of computational cost.

The observation of bidirectional transport in our simulations (Fig.~\ref{fig:spine_sim}(c)) is supported by experimental observations in \cite{Da2015} and previous studies of molecular motors present in dendritic spines. Restricting attention to actomyosin-based vesicle trafficking, which is thought to be the dominant mechanism for vesicle entry into spines \cite{Da2015}, we are aware of two plausible mechanisms for bidirectional transport. First, myosin types II, V, and VI are reported to be present in spines \cite{kneussel2013myosin}, and whereas all other known myosin species walk toward the barbed (+) end of actin, myosin VI walks toward the pointed (-) end \cite{wells1999myosin}. Second, actin of mixed polarities has been observed in spine necks \cite{kneussel2013myosin}, so that a single-directional myosin motor could move in either direction based on the actin bundle to which it is attached. 

%Because our model makes the quasi-steady state assumption that the vesicle motion is slow compared to all other physical processes, given multiple branches of solutions we face the potential issue of jumping from one branch to another between subsequent timesteps. In most practical situations, we find that using the previous solution as the initial guess for fixed point iteration gives the appropriate sequence of solutions. However, when there are multiple nearby fixed points, as is the case when the constriction is very narrow, and noise is added to the system, the solutions of fixed point iteration are observed to jump between branches. In particular, this may lead to cycling between unstable solutions. We find that ...   %One way to manage this issue is to relax the constraint $U = -Q/(\pi R_p^2)$ and to replace it by an evolution equation $\dot{U} = k(-Q/(\pi R_p^2)-U)$ for $U$, in which $k$ is a penalty parameter that recovers the constraint in the limit $k \to \infty$. This is equivalent to a simple elastic model of the pocket, with $k$ the elastic constant. We find that 

\section{Discussion}

By performing fully resolved 3D simulations using the lattice Boltzmann method, we have been able to study the translocation of an impermeable elastic vesicle into a closed tube filled with incompressible fluid. If the vesicle radius is nearly equal to the radius of the channel neck, the vesicle deforms into a teardrop-like shape and is separated from the channel walls by a thin lubrication layer. Because the fluid trapped in the pocket must escape, this fluid layer includes both forward and reverse flow, unlike the case of an open channel in which only forward flow is observed. This reverse flow decreases the size of the effective lubrication layer for the closed channel and increases the amount of force required for translocation at a given speed.

To capture the essential behavior, we have developed a lubrication model that involves only two nondimensional parameters involving the gap size and applied forcing and allows us to efficiently explore a large parameter range. We have found that the dependence of the transit time $\tau$ on minimum lubrication layer width $h_0$ scales as $ \tau \sim  h_0^{-5/2}$ in the inelastic regime, and these results have been validated by a combination of simulation and scaling analysis. We now make a brief digression to compare these results to those obtained in previous work on the forced transport of elastic containers through constrictions \cite{Kusters2014}. In that work, which focused on the case of large vesicles and identified a finite force under which the vesicle could not pass, the elastic properties of the vesicle dominated over fluid effects such as the formation of a lubrication layer. Here, the limit of interest is the one in which the vesicle size is nearly equal to the constriction size and lubrication effects play a crucial role, necessitating a more realistic model of fluid drag such as the the lubrication model we have used. Moreover, whereas in \cite{Kusters2014} the vesicle was assumed to pass through a series of known shapes, allowing for a computation of the energy and therefore the motor force required as functions of position through the channel, here we use a simple model for the vesicle shape changes (i.e.\ linear compliance). The lubrication assumption is invoked at all points through the vesicle transit, and the coupling of the shape to the pressure via the linear compliance results in a vesicle that slows down according to a power law but never stops moving through the channel. Combining this lubrication model with a more comprehensive series of allowed shapes would enable capturing the relevant fluid dynamics as well as the elastic barriers to translocation.

\edrs{Developing more realistic models of the vesicle's elastic response at large deformations that capture nonlinear stiffening effects would improve on the simple linear compliance model. Removing the assumption of small deformations would also allow us to explore the case $\pi_1 > 1$, which is of significant biological interest. As noted above, we believe the effects investigated here in the small deformations limit are applicable to the case of large deformations, and we expect the impeding effect of a closed end on the vesicle motion to become even more dramatic because of the narrower lubrication layer involved. We have not yet been able to investigate the case of large deformations since, in addition to the modeling challenges of accurately capturing vesicle shape and elasticity, it is difficult to resolve the smaller lubrication layers involved because of the maximum grid size allowed in the lattice Boltzmann simulations. Efforts are underway to overcome these challenges. In particular, some of us have begun developing a numerical method for fluid-structure interaction that combines a direct Navier-Stokes solver for the bulk flow with lubrication theory in regions with thin fluid layers \cite{lubeinprep}.}

We have added an additional level of realism to our model by including a mechanistic force-velocity relation for the forces exerted by molecular motors. The motor model we use includes two species with opposite polarities and exhibits multistable dynamics, consistent with previous studies in the literature of mixtures of molecular motors. Our work demonstrates that this multistability can interact with the channel geometry in fascinating ways, and our model is able to reproduce several of the behaviors observed experimentally in spines. The number and direction of steady state velocities can be controlled by varying the channel radius and the fraction of motors pushing in either direction, and these control mechanisms may be used by the cell to control dendritic spine maintenance, growth, and atrophy. \edrs{As mentioned above, vesicle elasticity does not play a critical role in Section \ref{sec:lube_motor} and therefore is not an essential ingredient for the multistable dynamics. Performing an in-depth parameter space investigation in the future would make it possible to see how the behavior changes in regions where fluid flow and molecular motors are strongly coupled to changes in vesicle shape.}

The present model makes several simplifying assumptions that must be refined for more specific quantitative models, e.g.\ biophysically detailed studies of vesicle trafficking into dendritic spines. For instance, we have left out the effect of fluctuations caused by Brownian motion and motor kinetics. The formulation of the lubrication model could be revisited to include thermal fluctuations, and the continuum theory of molecular motors used here could be replaced by a kinetic model to explicitly include the stochastic effects of motors coming on and off. Given the sub-micron scale of dendritic spine necks, we expect stochasticity to be an important ingredient in more quantitative studies of vesicle trafficking. Intracellular environments such as dendritic spines are crowded environments filled with various biomolecules, and the assumption of a surrounding Newtonian fluid could be improved by accounting for these non-Newtonian effects. Finally, the presence of a strong reverse flow through the lubricating layer depended crucially on the assumption of impermeability to water of both the spine wall and vesicle membrane. This is consistent with experimental evidence that mammalian neurons do not contain water-permeable membrane channels called aquaporins \cite{pyrneur}, but experimental investigation of the permeability of recycling endosomes is necessary to validate this assumption.

%\begin{figure}
%\centering
%\includegraphics[scale=0.7]{lube_model_page_new.pdf}
%\caption{\label{figure4}}
%\end{figure} 

%Note that, using \eqref{eqn:uit}, we could equivalently eliminate $U^i$ from \eqref{eqn:qit} and replace it by
%\begin{equation}
% this equation is incorrect as written
% Q^{i+1} = -\pi R^2 \Delta p\Bigg/\left(6 \mu \int_0^L \frac{1}{(h^i)^2(x)}\ud s + 2 \int_0^L \frac{1}{(h^i)^3(x)}\ud s\right). \label{eqn:qit}
%\end{equation}
%In practice, we have found the difference between iterating three and four equations to be negligible.

\section{Acknowledgments}
We thank the anonymous reviewers for their constructive feedback and assistance in improving this manuscript. The authors wish to acknowledge financial support under National Science Foundation grant DMS-1502851 (TGF), the Netherlands Organization for Scientific Research (NWO-FOM) within the program "Barriers in the Brain: the Molecular Physics of Learning and Memory" (No. FOM-E1012M), and the NSF grant DMR 14-20570 through the Harvard MRSEC (CR, LM), and the MacArthur Foundation (LM).

\appendix

\section{Determination of Compliance $C$}\label{app:comp}

The elasticity of the vesicle in our lubrication model is captured by a single-parameter, the compliance, which determines the deflection of the vesicle in response to the pressure. To be able to use the model to simulate vesicles with known material properties, we need to determine the appropriate compliance. To do so, we consider a spheroidal neo-Hookean surface parameterized by coordinates $\mb{q}=(q_1,q_2)$ with the deformed configuration $\mb{X}=\mb{X}(\mb{q})$ and reference configuration $\mb{Z}=\mb{Z}(\mb{q})$ in $\mathbb{R}^3$. Following the formulation described in \cite{vc_ib,Wu2015}, the elastic energy $E$ of such a surface with bulk modulus $\kappa_\alpha$ and shear modulus $\kappa_s$ is given by 
\[E = \frac{\kappa_\alpha}{2} \int_{S'} (\lambda_1\lambda_2-1)^2 \ud a' + \frac{\kappa_s}{2} \int_{S'} \left(\lambda_1/\lambda_2+\lambda_2/\lambda_1-2\right) \ud a', \]
where $\lambda_1$ and $\lambda_2$ are the principal stretch ratios, i.e.\ $\lambda_1^2$ and $\lambda_2^2$ are the eigenvalues of the $2 \times 2$ matrix $G_0^{-1}G$, where $G_{ij} = (\partial X_k/\partial q_i) ( \partial X_k/\partial q_j)$ and $(G_0)_{ij} = (\partial Z_k/\partial q_i )( \partial Z_k/\partial q_j)$, using the Einstein convention for summation over repeated indices. Let the reference configuration $\mb{Z}=(z_1,z_2,z_3)$ be given by the sphere with radius $R$ so that
\begin{align*}
 z_1 &= R\sin q_1\sin q_2 \\
  z_2 &= R\sin q_1\sin q_2 \\
   z_3 &= R\cos q_1
 \end{align*}
 where $q_1 \in [0,\pi)$ and $q_2 \in [0,2\pi)$ are the usual polar and azimuthal coordinates, respectively. Now, consider oblate spheroidal deformations $\mb{X}=(x_1,x_2,x_3)$ parametrized by
\begin{align*}
 x_1 &= R\sqrt{\sigma}\sin q_1\sin q_2 \\
  x_2 &= R\sqrt{\sigma}\sin q_1\sin q_2 \\
   x_3 &= R/\sigma\cos q_1
 \end{align*}
with $\sigma > 1$. (Taking $\sigma < 1$ would give a prolate spheroid.) Note that the axes are scaled so that the total volume enclosed by the surface is independent of $\sigma$. Direct computation gives
\[G_0 =  \begin{pmatrix}
1 & 0\\
0  &  R^2 \sin^2 q_1
\end{pmatrix},\,\,
G =  \begin{pmatrix}
R^2 \sigma \cos^2q_1+R^2/\sigma^2\sin^2q_1 & 0\\
0  &  R^2 \sigma \sin^2 q_1
\end{pmatrix}.
\]
Neglecting the poles which are a set of measure zero, we have therefore
\[G_0^{-1} G 
 =  \begin{pmatrix}
\sigma \cos^2q_1+1/\sigma^2\sin^2q_1 & 0\\
0  &  \sigma
\end{pmatrix},
\]
which implies that $\lambda_1^2\lambda_2^2 = \det (G_0^{-1} G) = \sigma^2\cos^2q_1+\sigma^{-1}\sin^2q_1 = \sigma^{-1}\left(1+(\sigma^3-1)\cos^2q_1\right)$. It follows that the elastic energy is
\begin{align}
\begin{split}
 E &= \frac{\kappa_\alpha}{2}\int_{q_1}\int_{q_2} (\lambda_1\lambda_2-1)^2 \det{G_0}^{1/2} \ud q_1 \ud q_2\\
&= \frac{\kappa_\alpha}{2}\int_{q_1}\int_{q_2} (\sigma^{-1/2}\left(1+(\sigma^3-1)\cos^2q_1\right)^{1/2}-1)^2 \det{G_0}^{1/2} \ud q_1 \ud q_2,
\end{split}
\end{align}
where we have neglected the shear energy and used the fact that $\det{G_0}^{1/2}$ is the area element in reference coordinates.

Letting $\sigma = 1+\epsilon$ for $\epsilon \ll 1$, $\sigma^{-1/2} \approx 1-\epsilon/2$ and $\sigma^3-1 \approx 3\epsilon$. Therefore
\begin{align}
 E &\approx \frac{\kappa_\alpha}{2}\int_{q_1}\int_{q_2} ((1-\epsilon/2)\left(1+3\epsilon\cos^2q_1\right)^{1/2}-1)^2 \det{G_0}^{1/2} \ud q_1 \ud q_2 \\
 &\approx \frac{\kappa_\alpha}{2}\int_{q_1}\int_{q_2} \left(-\frac{\epsilon}{2}+\frac{3\epsilon}{2}\cos^2q_1\right)^2 \det{G_0}^{1/2} \ud q_1 \ud q_2 \\
 &= \frac{\kappa_\alpha(a\epsilon)^2}{8}\int_{q_1}\int_{q_2} \left(-1+3\cos^2q_1\right)^2 \sin q_1 \ud q_1 \ud q_2,
\end{align}
where we have used $\det{G_0}^{1/2} = a^2 \sin q_1$ in the final equality. Evaluating the remaining integral is straightforward and results in a value of $16\pi/5$. Thus,
\begin{equation}
 E = \frac{2\pi \kappa_\alpha}{5}(a\epsilon)^2,
 \label{eqn:comp}
\end{equation}
so that the effective spring constant is $4\pi \kappa_\alpha/5$. Since the force $F$ on the oblate ellipsoid satisfies $F = (4\pi \kappa_\alpha/5) (a\epsilon)$ whereas the compliance is related to the pressure by $C p = a\epsilon$, with $p \approx F/(4 \pi R_p^2)$,
\[C \approx 4\pi R_p^2 (4\pi \kappa_\alpha/5)^{-1}.\]
Taking representative values of $R_p = 1$ \textmu m and $\kappa_\alpha = 1 \times 10^{-3}\, \text{N}/\text{m}$ yields an estimate for the compliance $C \approx 5 \times 10^{-9}\, \text{m}/\text{Pa}$. Note that we have kept only terms involving $\kappa_\alpha$ and have neglected the additional elastic moduli $\kappa_B$ and $\kappa_s$. The good agreement between the compliance derived in this appendix and the compliance estimated from the full 3D simulation justifies our assumption that the membrane's elastic properties are essentially determined by the single modulus $\kappa_\alpha$ in the regime of interest.

\section{Nondimensionalization}\label{app:nondim_lubem}
Here, we present a nondimensionalized system of equations for the lubrication model. As stated in Section \ref{sec:abs}, there are two nondimensional groups $\pi_1$ and $\pi_2$ that govern the fluid-structure interaction dynamics: $\pi_1 = R_p/R_c$, where $R_c$ is the radius at the narrowest point in the channel, and $\pi_2 =CF/(\pi R_p^3)$, where $C$ is the compliance and $F$ is the strength of forcing. Note that in the context of the 3D membrane model we may write $\pi_2 = 5F/(\pi \kappa_\alpha R_p)$ according to \eqref{eqn:comp}. Of course, a complete investigation of the 3D problem would require additional dimensionless groups, e.g.\ those involving the length of the channel \cite{Kusters2014} and the additional elastic moduli $\kappa_B$ and $\kappa_s$. However, as mentioned previously, in the lubrication limit the essential behavior is captured by representing the channel geometry with the single parameter $R_c$ and the vesicle elasticity with a single parameter $C$ based on the area dilation modulus $\kappa_\alpha$.

%The dimensionless group $\pi_3$ is special in that we are not free to choose its value--in free space $\pi_3 = 1$  according to the Stokes law, whereas in confined geometries its value depends on the minimal spacing $h$. 

We introduce the nondimensional variables $\widetilde{z} = z/R_c$, $\widetilde{h} = h/R_c$,  $\widetilde{U} = 6 \pi R_p \mu U /F$, $\widetilde{Q} = 6 \pi R_p \mu U /(F(\pi R_c)(R_c-R_p))$, and $\widetilde{p} = \pi R_p R_c p /F$. Rewriting the formulation \eqref{eqn:p}--\eqref{eqn:h} for the model under a fixed external force in nondimensional form, we have:
\begin{align}
 \widetilde{p}(\widetilde{z})-\widetilde{p}_0&=\widetilde{U}\int_{-\pi_1}^{\widetilde{z}} \widetilde{h}^{-2}(s)\ud s-\widetilde{Q}(1-\pi_1)\int_{-\pi_1}^{\widetilde{z}} \widetilde{h}^{-3}(s)\ud s \label{eqn:nonp} \\
  \widetilde{Q} &= (1-\pi_1)^{-1} \left(\widetilde{U}\int_{-\pi_1}^{\pi_1} \widetilde{h}^{-2}(s)\ud s -1\right)\Big/\left( \int_{-\pi_1}^{\pi_1} \widetilde{h}^{-3}(s)\ud s\right)\label{eqn:nonpL}\\ 
 \widetilde{U} &= -\pi_1 \widetilde{Q} \label{eqn:nonu}\\
 \widetilde{h}(\widetilde{z})&=1-\sqrt{\pi_1^2-\widetilde{z}^2}+\pi_2 (\widetilde{p}(\widetilde{z})-\widetilde{p}_0). \label{eqn:nonh}
\end{align}
Several interesting features of this model become readily apparent when written in the nondimensionalized form above. On the one hand, in the limit $\pi_1 \to 1$, the product $ \widetilde{Q}(1-\pi_1)$ tends to zero so that one might erroneously assume it can be neglected. In fact, it is precisely in this limit that $\widetilde{h} \to 0$, so that the integrand $\widetilde{h}^{-3}$ become singular and the term $\widetilde{Q}(1-\pi_1)\int_0^{\widetilde{z}} \widetilde{h}^{-3}(s)\ud s$ must be retained. Another property of this model is that the dimensionless variable $\widetilde{U}$ is not constant, as would be the case in free space according to Stokes' law. In confined geometries such as the channel considered here, its value depends on the nondimensional spacing $\widetilde{h}$.

Next, we consider how incorporating molecular motors into the lubrication model as in Section \ref{sec:lube_motor} affects the above nondimensionalized system. As described in \cite{hopp_pesk}, the motor model adds four nondimensional groups $\pi_i,\, i=3,\,4,\, 5, \, 6$, with $\pi_3 = \alpha/\beta$ the ratio of attachment and detachment rates, $\pi_4 = \gamma A$ the nondimensional attachment position, and $\pi_5 = \gamma (B-A)$ reflecting the maximum negative displacement of a motor. There is a fourth additional nondimensional group $\pi_6 = (F/ 6 \pi R_p \mu)/(\beta/\gamma)$ that gives the ratio of velocity scales between translocation and motor adhesion dynamics. Note that in the context of the motor model, the forcing constant to be used in $\pi_2$ is simply the stall force $F_0=\alpha p_1 n_0/(\alpha+\beta)(\exp(\gamma A)-1) $.

In terms of these nondimensional units, the motor model may be written
\begin{equation}
\widetilde{F}_A =
\begin{dcases}
      -\frac{1+\pi_6 \widetilde{U}\left(e^{\pi_4}-1\right)^{-1}}{1-\pi_6 \widetilde{U}},  & \widetilde{U} < 0\\
      -\frac{\pi_3+1}{\pi_3\left(1-e^{\pi_5/(\pi_6 \widetilde{U})}\right)}\frac{e^{\pi_4}\left(1-e^{\pi_5}e^{-\pi_5/(\pi_6  \widetilde{U})}\right)-\left(1-\pi_6  \widetilde{U}\right)\left(1-e^{-\pi_5/(\pi_6  \widetilde{U})}\right)}{\left(1-\pi_6  \widetilde{U}\right)\left(e^{\pi_4}-1\right)}, & \widetilde{U} \ge 0,
\end{dcases}
\end{equation}
where $\widetilde{F}_A = F_A/F_0$. The form of $F_{-A}$ is analogous:
\begin{equation}
\widetilde{F}_{-A} = 
\begin{dcases}
      \frac{\pi_3+1}{\pi_3\left(1-e^{\pi_5/(-\pi_6 \widetilde{U})}\right)}\frac{e^{\pi_4}\left(1-e^{\pi_5}e^{\pi_5/(\pi_6  \widetilde{U})}\right)-\left(1+\pi_6  \widetilde{U}\right)\left(1-e^{\pi_5/(\pi_6  \widetilde{U})}\right)}{\left(1+\pi_6  \widetilde{U}\right)\left(e^{\pi_4}-1\right)}, & \widetilde{U} < 0 \\
      \frac{1-\pi_6 \widetilde{U}\left(e^{\pi_4}-1\right)^{-1}}{1+\pi_6 \widetilde{U}},  & \widetilde{U} \ge 0.
\end{dcases}
\end{equation}
Setting $\widetilde{F} = \phi_1 \widetilde{F}_A+\phi_2 \widetilde{F}_{-A}$, the motors are then incorporated in the lubrication model by replacing \eqref{eqn:nonpL} by
\begin{align}
  \widetilde{Q} &= (1-\pi_1)^{-1} \left(\widetilde{U}\int_0^{\widetilde{L}} \widetilde{h}^{-2}(s)\ud s -\widetilde{F}\right)\Big/\left( \int_0^{\widetilde{L}} \widetilde{h}^{-3}(s)\ud s\right)\label{eqn:nonpLmm}
\end{align}

For our simulations in Section \ref{sec:compex}, we have used $\pi_3 = 1$, $\pi_4 = 4.7$, $\pi_5$ in the range $\pi_5 = 0.02$ -- $0.1$, and $\pi_6 = 10$ -- $18$. In terms of the key dimensional parameters, using a stall force of $F_0 = 50$--$200$ pN results in transit times on the order of $10$ -- $100$ seconds for vesicles of radius $R_p = 1$ -- $2$ \textmu m. This is consistent with the experimentally observed transit times of approximately $40$ -- $60$ seconds \cite{Da2015}.

\section{Lattice Bolzmann simulations}\label{app:LBsim}
\begin{figure}[h!]
\centering
\includegraphics[width=0.75\textwidth]{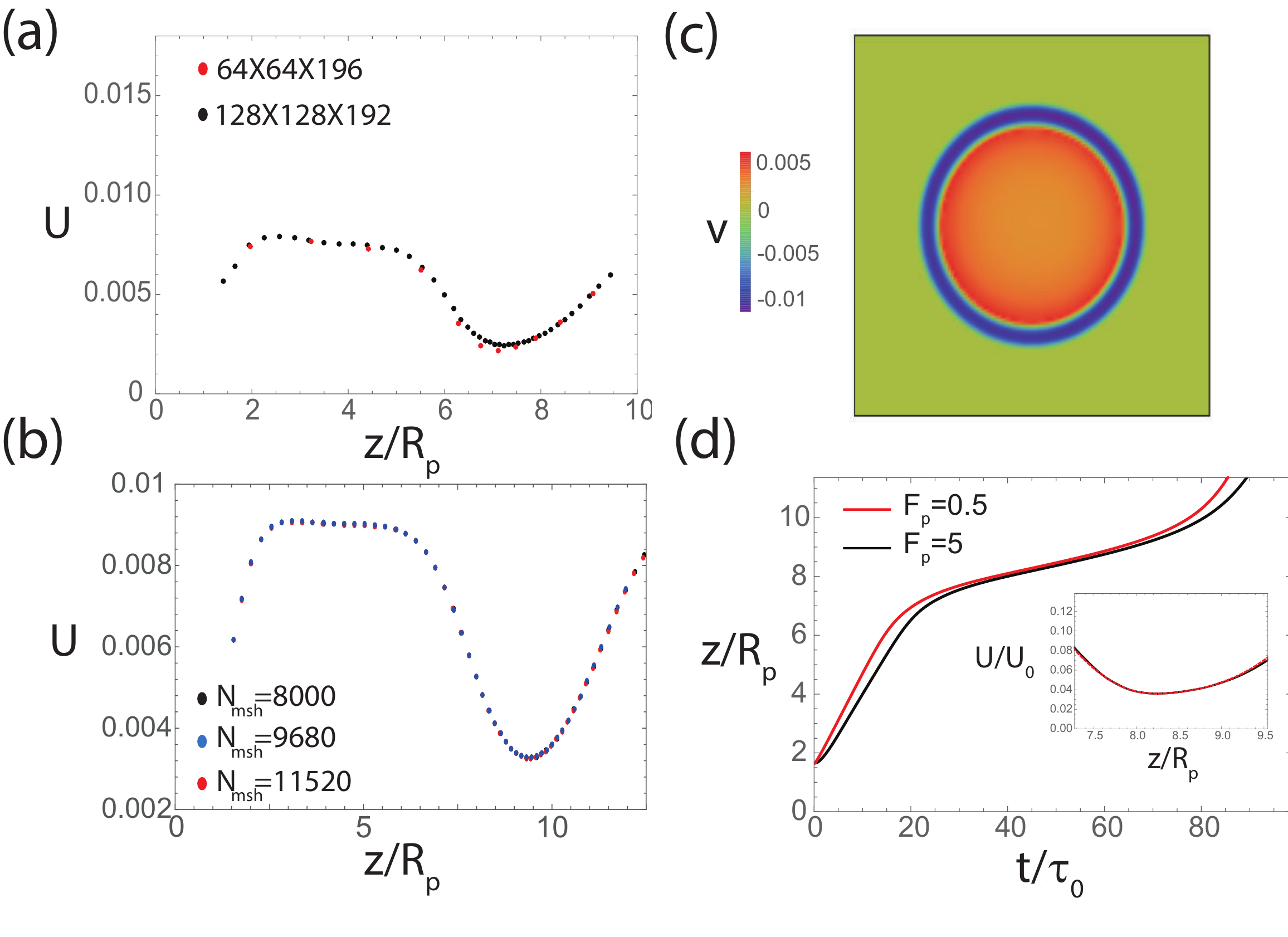}
\caption{ (a) Velocity of the vesicle as a function of position at two different grid resolutions and otherwise identical parameters. (b) Velocity of the vesicle as a function of position at three different mesh sizes. (c) Color-map of the $z$-velocity in the $x$-$y$ plane showing the radially symmetric velocity profile halfway through the constriction at $z=180$. \edrs{(d) Position of the center of mass of the vesicle as a function of time for identical dimensionless groups $\pi_1=0.73$ and $\pi_2=0.27$. Here the black curve corresponds to the moduli of Table \ref{tab:tableLB} and the red curve corresponds to elastic moduli and applied force one order of magnitude smaller. The inset shows that the resulting velocities in the constriction are nearly identical.}\label{SI1}}
\end{figure}

Here we discuss the convergence study that we have performed to validate the 3D simulation results obtained using the lattice Boltzmann method and show that the axisymmetric assumption of the lubrication model is supported by the 3D simulations. All the simulations in this paper have been performed on a $128\times128\times384$ grid. To ensure this resolution is sufficient to capture the narrow lubrication layer, we perform simulations using identical parameters on $64\times64\times192$ and $128\times128\times384$ grids ($\pi_1=0.62$ and $\pi_2=0.16$, and mechanical properties as in Table \ref{tab:tableLB}). In Fig. \ref{SI1}(a) we plot the velocity of the vesicle as a function of its position and find that the velocity profiles are nearly identical for both grids. This ensures that a system size of $128\times128\times384$ is sufficient to capture the translocation dynamics, at least for these system parameters. Similarly we have performed a convergence study on the mesh-size of the vesicle in Fig. \ref{SI1}(b) by calculating the velocity as a function of position for vesicles discretized using 8000, 9680 and 11520 mesh points. In this figure we show that all three mesh resolutions yield essentially the same trajectory. In all the simulations presented in this paper we have chosen the number of mesh points to be on the order of $480 R_c$, which has been shown to give converged results in the parameter regime used \cite{Krugerthesis, Kruger2011}.

The flow-field around the vesicle is axisymmetric during translocation, as illustrated by the radially-symmetric velocity profile of Fig. \ref{SI1}(c), which shows a color-map of the $z$-velocity in the $x$-$y$ plane halfway through the constriction at $z=180$ (using parameters as in Fig.~\ref{figure2}(a)). We have observed that this axisymmetric configuration is stable in the sense that axisymmetry is recovered after slight perturbations in the radial position of the vesicle.

\edrs{In Fig. \ref{SI1}(d) we compare the time evolution of the vesicle for identical dimensionless groups $\pi_1=0.73$ and $\pi_2=0.27$, where the black curve corresponds to the moduli as in Table \ref{tab:tableLB} and the red curve has an applied force and mechanical moduli that are one order of magnitude smaller ($F = 0.5$, $\kappa_{\alpha} = 0.1$, $\kappa_{s}=0.0015$ and $\kappa_{B}=0.0018$). Note that the rescaled velocity within the constriction is identical for both cases, ensuring that these dimensionless groups are the relevant groups to describe the translocation of the elastic vesicle through the constriction. }

\bibliographystyle{abbrv}                                        
\bibliography{biblio}

\end{document}